\documentclass[useAMS,usenatbib]{mn2e}
\usepackage[dvips]{graphicx}
\usepackage{txfonts}
\usepackage{amssymb}

\title[The dynamical formation of LMXBs in dense stellar environments]{The dynamical formation of LMXBs in dense stellar environments:
globular clusters and the inner bulge of M31}
\author[R. Voss and M. Gilfanov]{R. Voss$^{1}$\thanks{E-mail:
voss@mpa-garching.mpg.de (RV); gilfanov@mpa-garching.mpg.de (MG)} and
M. Gilfanov$^{1,2}$\footnotemark[1]\\
$^{1}$Max Planck Institut f\"ur Astrophysik, Karl-Schwarzschild-Str.1,
85741 Garching, Germany\\
$^{2}$Space Research Institute, Russian Academy of Sciences, Profsoyuznaya
84/32, 117997 Moscow, Russia}
\begin{document}

\date{}

\pagerange{\pageref{firstpage}--\pageref{lastpage}} \pubyear{2007}

\maketitle

\label{firstpage}

\begin{abstract}
The radial distribution of luminous ($L_X>10^{36}$ erg s$^{-1}$) X-ray
point sources in the bulge of M31 is investigated using archival
\textit{Chandra} observations. We find a significant increase of the 
specific frequency of X-ray sources, per unit stellar mass,
within 1 arcmin from the centre of the galaxy.
The radial distribution of surplus sources in this region follows the
$\rho_*^2$ law, suggesting that they are low-mass X-ray binaries
formed dynamically in the dense inner bulge.     
We investigate dynamical formation of LMXBs, paying particular
attention to the high velocity regime characteristic 
for galactic bulges, which has not been explored previously.
Our calculations suggest that the majority of the surplus sources
are formed in tidal captures of black holes by main sequence stars of
low mass, $M_*\la 0.3-0.4M_\odot$, with some contribution of NS systems of
same type. Due to the small size of the accretion discs a
fraction of such systems may be persistent X-ray sources. Some of
sources may be ultra-compact X-ray binaries with helium star/white dwarf
companions. We also predict a large number of faint transients, both NS and
BH systems, within $\sim$ 1 arcmin from the M31 galactic centre.
Finally, we consider the population of dynamically formed binaries in
Galactic globular clusters, emphasizing the differences between these
two types of stellar environments.

\end{abstract}

\begin{keywords}
galaxies: individual: M31 -- X-rays: binaries -- X-rays: galaxies 
\end{keywords}

\section{Introduction}
It is a well known fact that the ratio of the number of low mass
X-ray binaries (LMXBs)  to stellar mass is $\sim$ two orders of
magnitude higher in globular clusters (GCs) than in the Galactic  
disc \citep{Clark}. With the advent of \textit{Chandra} and
\textit{XMM-Newton}, studies of X-ray point sources in external
galaxies have become possible, and have shown that also there globular
clusters are especially abundant in LMXBs. This is attributed to
dynamical processes, through which LMXBs are formed in two-body
encounters. Due to the $\rho_*^2$ dependence on the stellar density  
such encounters are frequent in globular clusters and are negligible
in the field. Currently, there are 13 LMXBs \citep{Liu} in the
150 globular clusters  \citep{Harris} known in the Galaxy.  

In the central parts of massive galaxies, the stellar densities can
reach values similar to the densities in less luminous GCs. Except for
the very inner parts, these densities are still an order of magnitude
smaller than the densities found in the most 
luminous GCs, where the LMXBs are preferentially found. 
However, the large volume compensates for the smaller density
and LMXBs can be formed near the galactic centres
in two-body encounters in non-negligible numbers. 
Whereas dynamical interactions in globular clusters have been
intensively investigated, the parameter range typical of
galactic centres remains unexplored. 
Due to an order of magnitude higher stellar velocities, the character
of the dynamical interactions and relative importance of different
formation channels in the galactic centres differ from those in
globular clusters.  

Due to the large stellar mass contained in the central region of a
galaxy, a number of ``primordial'' LMXBs formed through the standard
evolutionary path exist there too. Although these can not be easily
distinguished from the binaries formed in two-body encounters, an  
argument of the specific LMXB frequency (per unit stellar mass) can be
employed, in the manner similar to the one that led to the discovery
of dynamical formation of binaries in globular clusters. 
The volume density of the primordial LMXBs follows the distribution of
the stellar mass in  a galaxy \citep{Gilfanov} whereas 
the spatial distribution of the dynamically formed binaries
is expected to obey the $\rho_*^2/v$ law \citep{Fabian}. 
Hence the latter should be expected to be much more concentrated
towards the centre of the host galaxy and reveal themselves as a
population of ``surplus'' sources near its centre.

M31 is the closest galaxy with a bulge density large enough to host a
number of LMXBs formed through dynamical interactions. At a distance 
of 780 kpc \citep{Stanek,Macri} X-ray sources can be easily resolved
with \textit{Chandra}, even near the centre of the galaxy. It
has   been studied extensively with \textit{Chandra} and we use these 
observations to explore the radial  distribution of bright X-ray
point sources in the bulge.  The results of this study are presented
in the Section \ref{sect:data} where it is demonstrated that the
specific frequency of X-rays sources increases sharply inside 
$\approx 1$ arcmin. 
The possible nature of surplus sources is discussed in section
\ref{sect:origin}. 
The details of dynamical formation of binaries in
dense stellar environments and dependence on the stellar velocity
dispersion are considered in the section \ref{sect:intro}.  
The results of this section are applied to the inner bulge of M31 and 
to the Galactic globular clusters in sections
\ref{sect:realistic}. Our conclusions are presented in the  
section \ref{sect:conclusions}.

\begin{figure}
\includegraphics[width=\hsize]{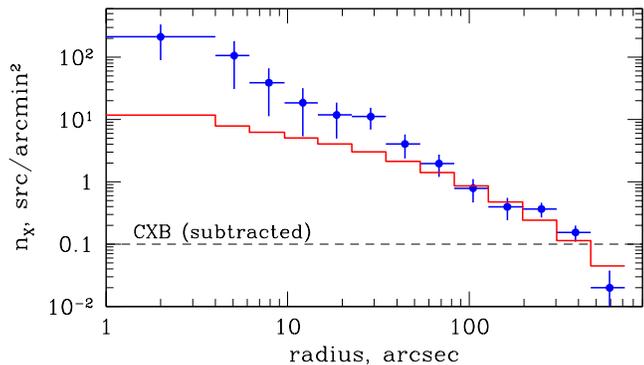}
\caption{The radial distribution of the X-ray point sources in
M31, excluding globular cluster sources and subtracting expected level
of CXB source density (shown by the dashed line). 
The histogram shows the distribution of primordial LMXB
sub-population as traced by the stellar mass distribution. 
The normalization of the latter is from the best fit
to the data outside 1 arcmin. 
}
\label{fig:fit}
\end{figure}

\section{Radial distribution of the X-ray point sources}
\label{sect:data}

With the currently available \textit{Chandra} data it is possible to 
study the spatial distribution of the X-ray point sources in the bulge,
without being affected by incompleteness, down to the limiting
lumionosity of 10$^{36}$ erg s$^{-1}$.  
We restrict our analysis out to a distance of 12 arcmin from
the centre and combine 26 ACIS observations with telescope pointings
within the central 10 arcmin region of the M31 bulge for a total
exposure time of 201 ks. Details of the data analysis, the source 
lists and the luminosity functions of various sub-populations in the
bulge are presented in \citet{Voss2}.

We model the radial distribution of the X-ray sources
by a superposition of primordial LMXBs and CXB sources, as in
\citet{Voss}. The spatial distribution of the former is assumed to
follow the stellar mass distribution of the galaxy, as traced by the
K-band light \citep{Gilfanov}. We used the K-band
image of M31 provided by 2MASS LGA \citep{Jarret}.
The distribution of  CXB sources is assumed to be
flat on the angular scales of interest.
Before proceeding with the fit we removed the contribution the from sources
other than primordial LMXBs and background galaxies. Firstly, we
removed 4 identified foreground sources, 1  supernova remnant and
one extended source. Secondly, we excluded X-ray binaries associated 
with globular clusters, as their origin and spatial distribution
are different from the ``field'' LMXBs.

Among our X-ray sources 13 are coincident with confirmed GCs from
\citet{Bologna} and 8 with GC candidates. We estimated the number 
of random matches by displacing the sources by 10 arcsec in 4 
directions. We found an average of 0.25 coincidences with confirmed 
GCs and 1.0 with GC candidates. 
It is well known that the inner parts of M31 are depleted of GCs.
\citet{Barmby} estimate that 70 per cent of the GCs within 5 arcmin
from the centre of M31 have been detected, leaving 
$\sim$16 GCs undetected. As only a fraction,
$\sim 1/5$, of the GCs in M31 contain LMXBs the expected contribution
of LMXBs from undetected GCs is $\sim$3. Due to selection
effects, the  majority of the undetected GCs are of low luminosity
(absolute visual magnitudes $\textit{V}\gtrsim-7$), and as LMXBs are 
preferentially found in high luminosity GCs the actual number of
LMXBs in undetected GCs is expected to be $\lesssim 1$.
A large fraction of the GC candidates are not real globular
clusters. However, an association with an X-ray source raises the  
probability of the GC candidates actually being GCs considerably. We
therefore remove these sources from our source list too, noting that
all conclusions  of this paper remain unchanged if the analysis is
performed with a source list in which these sources are included. 

We fit the relative normalizations of the LMXBs and CXBs, using the 
maximum likelihood (ML) test. The best fit is given by a model,
in which the normalization of CXBs is zero, meaning that all the
sources are LMXBs. As an alternative, we performed a $\chi^2$-fit
on the binned data, with $>15$ sources in each bin, 
and obtained the same result.
The probability that the data can be a realization of the model
is 0.06 using the Kolmogorov-Smirnov (KS) test, and $6\cdot 10^{-4}$  
for the $\chi^2$
test. The KS test is less sensitive to deviations at the end 
of a distribution, and therefore the result of the
$\chi^2$-test is more restrictive. We conclude that
the LMXB+CXB model is rejected.

The visual examination of the data (Fig. \ref{fig:fit}) suggests
that the reason for the rejection of the model is an overdensity of
sources in the inner 1 arcmin region of M31. Motivated by this we did
a $\chi^2$ fit of the same model to the distribution outside 1
arcmin. The best fit value of the normalization of the CXB component 
gives the total number of 26$\pm$9 sources CXB sources in the entire 
$r<12$ arcmin.
This value is consistent with the expectation
of 29 background galaxies, estimated from the soft band of
\citet{Moretti}, using the method described in \citet{Voss}. 
We therefore fix the normalization of the CXB component at the
value corresponding to 29 sources.
This gives a total number of the LMXBs of 64$\pm$7 in the entire 
$r<12$ arcmin image. The $\chi^2$-value is 2.63 for 3 degrees of
freedom. The best fit model is shown in Fig. \ref{fig:fit} together
with the observed distribution. 

Using the best-fit model it is possible to investigate the
distribution of  sources in the inner 1 arcmin and quantify the excess
in the surface density of the sources. 
The total number of sources detected in the the $r<60$ arcsec
region is 29. The extrapolation of the best  fit model into this
region predicts 8.4$\pm0.9$ sources, and therefore the number of
surplus sources is 20.6$\pm5.5$. The error in the latter 
estimate accounts for the Poissonian uncertainty in the total number
of sources inside 60 arcsec and for the uncertainty of the best fit
model normalization. As it is obvious from Fig.\ref{fig:fit}, the
contrast between the observed number of sources and that predicted
from the K-band light distribution increases towards the centre of the
galaxy. Inside $r<15$ arcsec, for example, 9 sources are detected with
only 1.1 sources predicted. The formal
probability of such an excess to happen due to statistical fluctuation
is  $\sim 3\cdot 10^{-6}$, assuming Poissonian distribution.

\section{Origin of the surplus binaries}
\label{sect:origin}

Non-uniform extinction, peaking at the centre of M31, could cause the 
distribution of the \textit{K}-band light to deviate from the
distribution of stellar mass. This possibility can be excluded,
however, as the extinction towards the centre of M31 is low,
$A_V$=0.24 mag and $A_I$=0.14 mag \citep{Han}, which extrapolated
to the \textit{K}-band gives $A_K$=0.03 \citep{Binney}. Moreover, a 
non-uniform 
extinction distribution would also  cause non-uniformity in the
apparent colours of the stellar population, which is not observed
\citep{Walterbos}.  

The surplus sources can be high-mass X-ray binaries associated with
star formation in the inner bulge of M31. 
We derive upper limits for the star formation rate and the number of
HMXBs from the H$\alpha$ and FIR luminosities reported by 
\citet{Devereux}. The combined H$\alpha$ luminosity  of
the nuclear region and from diffuse emission inside the star
forming ring (which lies at a radius $\sim$50 arcmin, i.e.
much larger than the region analysed in this paper) 
is 4.3$\cdot$ 10$^{39}$  erg s$^{-1}$ (transformed to the distance of
780 kpc used in this paper). From \citet{Grimm} we find that this
corresponds to 3.2 HMXBs with a luminosity above 10$^{36}$ erg
s$^{-1}$. The FIR luminosity in this region is 5.25$\cdot$10$^{8}$
L$_\odot$, which corresponds to 5.9 HMXBs with a luminosity above
10$^{36}$ erg s$^{-1}$. 

It should be stressed out, that the region these luminosities
refer to is almost 20 times larger than the region analysed in this
paper. Moreover it is very likely that the main part of 
the H$\alpha$ and FIR emission is not associated with star formation,
as  the number of O-type stars is a factor of $\sim$ 200 lower than
what would be expected otherwise \citep{Devereux}.   
To conclude, the HMXB nature of the sources in the inner bulge can be
excluded.

The surplus sources could have been created in globular clusters that
remain undetected. 
In the catalogue of \citet{Bologna} there are 64 confirmed GCs
hosting 13 LMXBs in the region we analysed. The fraction of GCs
containing X-ray sources is therefore 0.2. This number is larger than
what is found in 
other galaxies \citep{Sarazin,Maccarone}, due to the better sensitivity
of our study, but consistent with
the results for the inner parts of galaxies in
\citet{Kim}. Attributing the $\approx 20$ surplus sources to
undetected GCs would therefore indicate that $\sim$ 100 unobserved GCs
exist in the inner 1 arcmin region of M31. 
This is much larger than allowed by the completeness level of the
present studies of GC population in M31, consistent with 
only a few undetected globular clusters in this region
\citep{Barmby}. 

\begin{figure}
\includegraphics[angle=270]{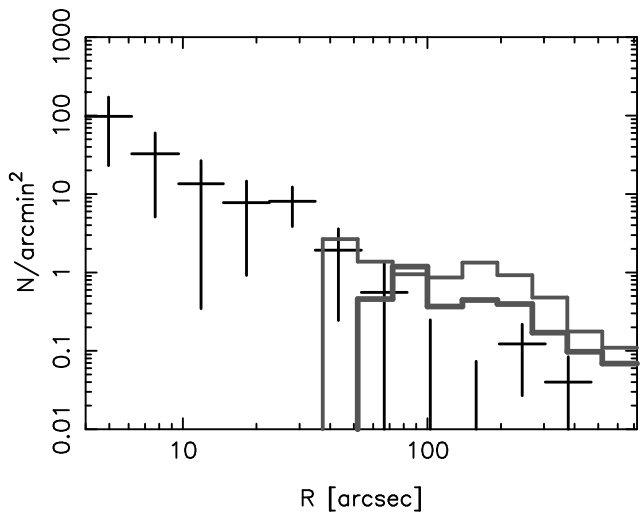}
\caption{The radial distribution of the ``surplus'' X-ray sources, computed
  as a difference between the data and best fit model shown in figure
  \ref{fig:fit}. The histograms show distributions of the confirmed
  globular clusters (thick grey line) and globular cluster candidates
  (thin grey line).} 
\label{fig:surplus2}
\end{figure}

In a related scenario, the surplus sources could have been created in 
globular clusters at larger distances from the centre of M31. 
Due to the mass segregation the globular clusters drift towards the
centre of the bulge, where they are destroyed, leaving behind remnant
LMXBs. This scenario has been motivated by Fig.\ref{fig:surplus2}
where the radial distribution of surplus sources is compared with that
of the globular clusters.
Indeed, for a GC of mass $10^5 M_{\odot}$ the mass segregation 
timescale is $\sim 10^9$ yr at a radius of 5 arcmin and
 $\sim 10^{10}$ yr at a radius of 12 arcmin
\citep{Spitzer}. 
Assuming that a neutron star turns accreted matter into
radiation with an efficiency of $\sim$ 0.2, the lifetime of an
LMXB is $\la 10^9 m_d/L_{37}$ yr, where $m_d$ is the mass of the donor
star at the onset of mass transfer expressed in solar masses, and
$L_{37}$ is the average luminosity of an LMXB in units of $10^{37}$
erg s$^{-1}$.  Taking into account that on average $\sim1/5$ of GCs in
M31 contain LMXBs, a destruction rate of $\sim$ 100 globular clusters per
Gyr is required to explain $\approx 20$ sources observed near the
centre of M31. This number is comparable to the total number of GCs
within the entire region analysed in this study, and is $\sim$ 30 per
cent of the total number of GCs in M31. As GCs are not continually
formed in large numbers in M31, the  globular cluster system of this
galaxy will not be able to sustain such a destruction  rate and, 
consequently, the population of X-ray sources observed in the inner
bulge,  for any  significant length of time.

Finally, the stellar density in the central part of the M31 bulge,
$\sim 10^4$ M$_\odot$/pc$^3$, is high enough that LMXBs can be formed
through dynamical interactions in the same manner as in globular
clusters.  In the following sections we investigate this possibility, 
and apply it to the population of X-ray sources in the inner bulge of
M31 and in globular clusters in the Milky Way.

\section{Dynamical interactions in dense stellar environments}
\label{sect:intro}

There are three main channels of dynamical LMXB formation operating in
dense stellar environments\footnote{We begin with consideration of 
encounters involving a neutron star. These results are
extended to formation of black hole binaries in the section
\ref{sect:bh}.}: 
\begin{enumerate}
\item In a tidal capture of a neutron star
(NS) by a non-degenerate single star, a close passage of the two stars 
induces oscillations in the non-degenerate star, and the energy
for this is taken from the orbital energy. If the energy of the
oscillations exceeds the originally positive orbital energy, the
stars are captured in a bound orbit \citep{Fabian}. 
\item A collision between an NS and an evolved single star on the
subgiant or red giant branch (RGB) or the asymptotic giant branch
(AGB) can lead to the formation of an X-ray binary,
in which the donor star is a white/brown dwarf or a helium star,
depending on the evolutionary stage of the evolved star before the
collision \citep{IvanovaU}. In the case of a white dwarf donor an
ultra compact X-ray binary is formed. In this scenario, orbital energy  
is transferred to the envelope of the evolved star, which is
expelled, leaving the NS and the core of the evolved
star in a bound orbit \citep{Verbunt2}. 
\item In an exchange reaction,
an NS exchanges place with a star in a pre-existing
binary during a close binary-single encounter \citep{Hills}. 
\end{enumerate}

In the context of LMXB formation in globular clusters in the Milky
Way, the  attention has been initially drawn to the tidal captures
\citep{Fabian}, while the potential importance of the two other
mechanisms has been realized few years later \citep{Hills,Verbunt2}. 
The estimates of the LMXB production rates, which followed, revealed
that each of the channels could give significant contribution to the
population of LMXBs found in the Galactic GCs \citep[e.g.][]{VerbuntH,DaviesB}. 
Since then large amount of work has been done  
to understand physics of stellar encounters in detail, explore
their parameter space and derive accurate mathematical
prescriptions for the crossections and rates \citep[e.g.][]{Press,Lee,
McMillan,Rasio,Davies,Sigurdsson,Heggie3}.
However, there have been surprisingly few studies making
specific predictions of numbers of dynamically formed binaries which
could be directly compared with their observed 
population in Galactic GCs.
With the exception of a few studies considering a handful of
individual GCs \citep{DaviesB,IvanovaU}, 
the rates are usualy computed for a set of the 
representative values of parameters (such as stellar density, velocity
dispersion etc.) and then extrapolated to the entire Galactic globular 
cluster system \citep[e.g.][]{Verbunt2}. An often used assumption is also that the
number of LMXBs is proportional to $\rho_{\ast}^2/\sigma_v$
\citep{VerbuntH,Pooley}. Although acceptable as an initial approximation, it is too
crude to perform a quantitative comparison of the theory with
observations.  Another major limitation of the most of these
investigations is that the subsequent evolution of the newly formed
binary, into and through the X-ray active phase, is ignored. 
Due to lack of the attention and effort in this
direction, it is currently unclear if any of the channels strongly
dominates over the others in real globular clusters. Even less 
understood is the  operation of these processes in the environment of
the galactic centers.  

It is the goal of this paper to fill these gaps.
In particular, a special attention will be paid to the following
aspects of the problem, which have often been
ignored in the previous publications on this subject: 
\begin{enumerate}
\item calculation of the encounter crossections and
rates in the high velocity regime typical for the galactic centers and
investigations of their velocity dependence
\item critical review and comparison of the stellar environments
(present day mass function, metallicity, abundance of compact objects
etc) in globular clusters and galactic centers and investigations of
their impact on the overall LMXB production rates via
different LMXB formation channels. Reasonably accurate calculation of
the encounter rates in the Galactic globular clusters and inner bulge
of M31, based on their structural properties. 
\item account for evolution of the newly formed binary before and
during the X-ray active phase and estimates of the expected numbers of
LMXBs based on the derived encounter rates
\end{enumerate}

\subsection{General considerations}

In the following we compute cross-sections and rates
of the three formation channels, consider their dependence on the velocity
dispersion of the stars and discuss various factors
affecting their efficiency in the high velocity regime.

In section \ref{sect:realistic} we use these results to calculate
theoretical formation rates and numbers of observable LMXBs in the
bulge of M31 and in the Galactic GCs.

Each of the processes depends on the rate of encounters between
two types of objects, which in a unit volume is given by
$n_1n_2\gamma$, where
\begin{equation}
\label{eq:rate1}
\gamma=\int_0^\infty F(v_{rel},\sigma_v)\sigma(v_{rel},M_1,M_2) v_{rel}dv_{rel}
\end{equation}
where $n_1$ and $n_2$ are the number densities and $M_1$ and $M_2$ are the
masses of object type 1
and 2, respectively, $\sigma (v_{rel})$ is the cross-section of the
encounter, and $F(v_{rel})$ is the distribution of relative velocities
at infinity. Assuming that the
velocity distributions of the two kinds of objects are both Maxwellian
and have the same three-dimensional velocity dispersion $\sigma_v$,
the distribution of  
relative velocities is given by
\begin{equation}
\label{eq:veldis}
F(v_{rel})dv_{rel}=\left(\frac{4\pi}{3}\right)^{-3/2}\sigma_v^{-3}
exp\left(-\frac{3v_{rel}^2}{4\sigma_v^2}\right)4\pi v_{rel}^2dv_{rel}
\end{equation}
Due to the effect of gravitational focusing, the cross-section for
two objects to pass within a distance $D$ of each other is
given by
\begin{equation}
\label{eq:sigma}
\sigma (v_{rel})=\pi D^2\left( 1+\frac{2G(M_1+M_2)}{Dv_{rel}^2}\right)
\end{equation}
In most cases,  the gravitational focusing (the second term in the
brackets)  dominates. 
Only for very fast encounters (large  $D$ and/or  $v_{rel}$)
 is $Dv_{rel}^2 > 2G(M_1+M_2)$.
If $D$ is independent on the relative velocity, 
 $\gamma \propto \rho^2/v_{rel}$ for slow encounters,
and $\gamma \propto \rho^2v_{rel}$ for the fast ones.

Several remarks are in place, concerning the subsequent evolution of the
newly created binary system with a compact object.

Capture of a neutron star in a bound orbit with a companion will lead to
formation of an X-ray binary provided that the companion star will
fill its Roche lobe and mass transfer will commense within a
reasonable time, shorter than $\sim 5-10$ Gyr.  
If the initial binary separation is too large for this to occur
immediately after the capture, it  can be decreased in the course of 
evolution of the 
binary. There are 3 main mechanisms, which affect the orbital
separation: (i) magnetic braking, (ii) gravitational braking and
(iii) binary-single interactions.  The former two are familiar from the
standard theory of the binary evolution \citep[see][for a review]{Heuvel}. 
For the companion mass in the $0.3-1.0 M_{\odot}$ range they will bring the 
system in to the Roche lobe
contact within 5 Gyr if the initial orbital separation does not exceed
$\sim 3.0-7.0 R_{\odot}$ and $\sim 2.5-3.0 R_{\odot}$ respectively. 
The braking mechanism due to interaction of the binary with single
``field'' stars is specific for high stellar density environments.
Its properties are briefly summarized below.  

When considering evolution of a binary due to binary-single
interactions it is conventional to divide the binaries into soft and
hard, depending on the ratio of their binding energy to the kinetic
energy of the single star at infinity \citep{Heggie1}. 
Soft binaries have  relatively wide orbits, and interactions with
single stars tend to widen the orbit further or to ionize the
binary. Hard binaries, on the contrary, are on average hardened
by encounters with single stars \citep{Hut}.
The effect of this is that over time most binaries with
a separation above a critical value are disrupted, while the compact
ones become more compact. 
The boundary between soft and hard binaries depends on the stellar 
velocity dispersion and the mass ratios and ranges from $a\sim
300-1000 R_{\odot}$ in a typical globular cluster to $a\sim$ few
$R_{\odot}$ in the high velocity environment of the M31 bulge. 
Due to a linear dependence of the crossection on the binary separation
the collisional braking is mostly important at wide binaries,
where magnetic braking and gravitational radiation, decreasing as
inverse power of the binary separation, are  inefficient.  

The initial orbital separation in the  binaries produced through tidal
captures and  collisions with RGB/AGB stars is small and a large
fraction of this systems will start mass transfer (i.e. become X-ray
sources) soon after their formation.  
Only a small fraction of them ($\lesssim$20 per cent in GCs and $\lesssim2$
per cent in M31) will experience close encounters with single stars
significantly affecting thery semimajor axis, 
therefore binary-single interactions are not an important factor in
their evolution. 
Binaries created through exchanges, on the contrary, typically have
wider orbits, and the effects of encounters can be important.

If the initial binary separation is large and the braking mechanisms
are insufficient to start Roche-lobe overflow, this can occur when
the donor star evolves off the main sequence, as a result of its
expansion during the giant phase.  
In these systems the accretion disc is large and X-ray emission
from vicinity of the NS is insufficient for the irradiation to keep
the entire disc ionized, and they are therefore
transient  \citep{King2}. 
Furthermore, mass transfer can only occur while the donor is on the 
RGB, which makes the lifetime of such systems short. 
While they may account for bright sources detected
in massive elliptical galaxies \citep{Piro}, they are too rare 
to make a significant contribution to our sample.
The NSs in these systems are spun up to become millisecond
pulsars, and in the Galactic GCs a large number of these
have been observed \citep{Lorimer}. After the outer layers
of the giant star have been ejected, a binary consisting
of a white dwarf and an NS remains.
However for the vast majority of the systems the binary separation
is too large for mass transfer to begin, and they will
therefore not become observable in X-rays.

\subsection{Single-single encounters}
The formation rates of LMXBs due to tidal captures and
stellar collisions can be found by integrating the
encounter rate, given by equation \ref{eq:rate1}, over the relevant parts of
the mass function of stars $f(M)$. We assume that the latter follows 
the form of \citet{Kroupa}, a broken powerlaw with slope 1.3 from 
0.1-0.5$M_{\odot}$ and slope 2.3 above 0.5 $M_{\odot}$, 
and is normalized according to
\begin{equation}
\int_{M_{co}}^{M_{max}}f(M)dM=1.0
\end{equation}
where $M_{max}$ is the maximum initial mass of stars that have not
yet evolved to become stellar remnants at present, and $M_{co}$ is
the lower cut-off mass.
The number density of stars is then
given by $n_{\ast}=\frac{\rho_{\ast}}{<M>}$, where $\rho_{\ast}$ is
the stellar mass density. We assume the mass of all
neutron stars to be $1.4M_{\odot}$, and that they are formed
from stars with initial mass in the range $8.0-30.0M_{\odot}$. 
The number density of these can then be expressed as 
$n_{ns}=f_{ns}\frac{\rho_{\ast}}{<M>}=f_{ns}n_{\ast}$, where 
$f_{ns}=\int_{8M_{\odot}}^{30M_{\odot}}f(M)dM$ (for $M_{co}=0.1 M_{\odot}$
and $M_{max}=1.0 M_{\odot}$ $f_{ns}=0.0068$). 
We define the rate integrated with
the mass function
\begin{equation}
\Gamma=\int\gamma f(M)dM
\label{eq:gamma_large}
\end{equation}
where integration is performed in the relevant initial mass range (see below)
and $\gamma$ is from equation \ref{eq:rate1}.
With this definition $n_{\ast}n_{ns}\Gamma$ gives the rate of
encounters in s$^{-1}$ cm$^{-3}$.
For the calculation of $\Gamma$ it is necessary to know the current
radius of a star $R(M)$ as a function of its initial mass,
as well as its evolutionary stage, which is used to define the
mass limits of the integral. These informations we take from
stellar isochrones of \citet{Girardi}.

\subsubsection{Collisions}
\label{sect:coll}
We define an encounter between two stars as a collision if the stars
come so close that considerable amounts of material is exchanged between
them, and hydrodynamical effects become important. 
For the collisions with NSs, that are relevant
for dynamical formation of LMXBs, we distinguish between collisions
with main sequence (MS) or horizontal branch (HB) 
stars and with evolved stars on the RGB or AGB.
Due to the different structure of the stars, the outcome of a collision 
with a NS is different. Simulations indicate that collisions between
an NS and an MS star tend to destroy the MS star \citep{Davies}.
We expect the same to happen to stars on the HB,
as their structure is similar to that of MS stars \citep{Dorman}.
Collisions of this type are not interesting from the point of view of
formation of X-ray binaries.
As the envelope of stars on the RGB or AGB is less strongly bound
to the core, a collision with an NS can lead to the envelope
being expelled. The outcome is a short period binary consisting of
the core of the evolved star and the NS. If the evolved
star had a degenerate core, an ultra-compact X-ray binary (UCXB) with
a white dwarf donor will be formed \citep{Verbunt2,IvanovaU,Lombardi}.
In either case an X-ray binary can be created.

The maximum value of distance at periastron $R_{coll}$, for which 
significant amounts of material can be exchanged in an encounter
of an NS with a non-degenerate star (with radius $R_{\ast}$)
is between $R_{\ast}$ and the orbital separation at which the star fills its
Roche-lobe \citep{Eggleton}:
\begin{equation}
\label{eq:afill}
a_{fill}=\frac{R_{\ast}}{0.49}\left[0.6+q^{-2/3}\ln \left( 1+ q^{1/3}\right)\right]
\end{equation}
where $q=M_{\ast}/M_{NS}$. For
encounters with NSs, this separation ranges from 
$\sim 5.4 R_{\ast}$ to $\sim 2.8 R_{\ast}$ for stars with masses in the
$0.1-1.0 M_{\odot}$ range.
SPH Simulations of stellar encounters have shown
that for $M_{\ast} \simeq 1 M_{\odot}$, the value of $R_{coll}/R_{\ast}
\sim 1.8$ \citep{Davies}, 
which we adopt as the standard value. The value of $R_{coll}/R_{\ast}$
given, the encounter rate can be calculated from equations 
\ref{eq:rate1}-\ref{eq:sigma} and \ref{eq:gamma_large} (but see below
regarding the choice of integration limits in equation \ref{eq:rate1}).

When considering collisions between NSs and evolved stars, it
is important to note that the envelope of a star on the RGB/AGB is  
loosely bound to the core, and the orbital energy of the two stars at
infinity can be  comparable  to the binding energy of the
envelope. It is therefore possible 
that the envelope is expelled without carrying off enough energy
to leave a bound system, or that in the high-velocity encounters,
the duration of the interaction is
too short for enough energy to be transferred from the NS to the
envelope.   
While simulations indicate that the RG envelope is promptly disrupted,
instead of ejected through a common envelope (CE) evolution
\citep{Rasio,Lombardi}, the energy
considerations are similar.
Adopting the formalism
of \citet{Webbink} and \citet{de Kool}, we assume that the envelope
of the RG is ejected, and that energy for this (the binding energy
of the envelope $E_{bind}$) 
is taken from the orbital energy of the two stars which
is therefore changed by $\Delta E_{orb}$. Allowing energy to be
lost, e.g. as radiation or as some of the envelope is ejected with
a velocity higher than the escape speed, an efficiency parameter
$\alpha_{ce}$ is defined, so that $E_{bind}=\alpha_{ce}\Delta E_{orb}$. 
The binding energy of the RG is given by $E_{bind}=-\frac{1.0}{\lambda}
\frac{GM_{env}M}{R}$, where $M$ and $R$ is the mass and radius of 
the RG, respectively,
and $M_{env}$ is the mass of the envelope of the RG. $\lambda$
is a factor that relates the simplified equation to a precise
integral of the gravitational binding energy and internal
energy in the envelope of the RG, see e.g. \citet{Dewi}.
The change in orbital energy
needed to reach an orbit with a separation $a_f$ is given
by $\Delta E_{orb}=\frac{1}{2}\frac{MM_{ns}}{M+M_{ns}}
v_{rel}^2+\frac{GM_{core}M_{ns}}{2a_f}$ , where $M_{core}$ is the 
core mass of the RG, $M_{ns}$ is the mass of the NS and 
$v_{rel}$ is the relative velocity of the two stars at infinity.

\begin{figure}
\begin{center}
\resizebox{\hsize}{!}{\includegraphics[angle=270]{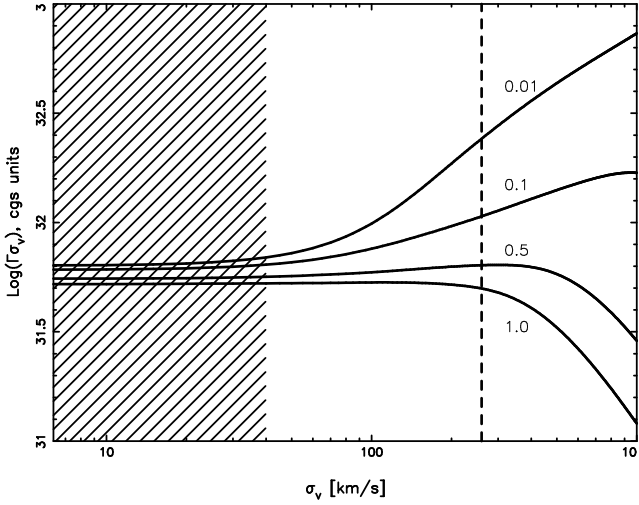}}
\caption{The rate $\gamma$ of NS-RGB/AGB encounters that lead
to the formation of a binary as a function of the 3D velocity dispersion
for 4 values of $\alpha_{ce}\lambda$ (equation \ref{eq:CE}). 
The values are 0.01,0.1,0.5,1.0 from the top to the bottom. The shaded
area corresponds to the range of  
velocity dispersions in the central parts of GCs, wheras the vertical
dashed line corresponds to the velocity dispersion in the bulge
of M31.}
\label{fig:vrel}
\end{center}
\end{figure}

For a given encounter velocity, we can now find the final separation
of the binary by solving
\begin{equation}
\label{eq:CE}
\frac{GM_{env}M}{R}=\alpha_{ce}\lambda\left(\frac{1}{2}\frac{MM_{ns}}{M+M_{ns}}v_{rel}^2+\frac{GM_{core}M_{ns}}
{2a_f}\right)
\end{equation}
When we calculate the rate of collisions with evolved stars, $\gamma_{coll}$,
the integral over velocities (equation  \ref{eq:rate1}) is only
carried out for velocities  
$V_{rel}<V_{max,c}$, defined such that the final separation
$a_f<5R_{\odot}$ (this choice of the maximum separation ensures that
the gravitational braking will be efficient on the formed binary)
\begin{equation}
\label{eq:coll}
\gamma_{coll}=\int_0^{v_{max,c}}F(v_{rel},\sigma_v)\sigma_{coll}v_{rel}dv_{rel}
\end{equation}
where $\sigma_{coll}$ is the collisional cross-section defined by
equation \ref{eq:sigma} with $D=R_{coll}$.
In figure \ref{fig:vrel} we compare the formation rates of UCXBs due
to collisions between RGB+AGB stars and NSs for different values of
$\alpha_{ce}\lambda$.  
The rates were calculated by integrating over all evolutionary
stages of stars on the RGB/AGB, using the isochrones of \citep{Girardi}.
The details of the calculations are given below in
section \ref{sect:comparison}. It is obvious that the choice of
$\alpha_{ce}\lambda$ is very important in the bulge of M31, with an order
of magnitude difference between the rates of the highest and the lowest
value, while the difference is relatively small in GCs.

SPH simulations indicate that the effectivity in standard CE 
evolution is $\alpha_{ce} \lesssim 0.5$
\citep[see][and references therein]{Taam}, and
in population synthesis studies, values of $\lambda\alpha_{ce}$ in the
range 0.1-1.0 are most often assumed
\citep{Portegies,Fryer,Hurley,VossT,Belczynski}, 
and this seems to give a good
fit to the observed properties of the post-CE binary
population.

However there are differences between the standard CE
evolution and the collisions considered here. In the former,
the stars are already in a bound orbit, and the energy can
therefore be transferred to the envelope over a longer period
of time during a large number of orbital revolutions. On the
other hand, in a collision, enough energy has to be transferred
from the NS to the envelope during the first periastron passage,
so that the two stars remain bound. Especially for high velocity
encounters, the CE formalism might not be directly applicable, as
the timescale for the first passage can be so short that the NS
passes through the envelope without transferring much energy to
this.
While the low-velocity regime has been well investigated using
SPH simulations \citep{Davies,Lombardi}, showing that a value
of $R_{coll}/R_{\ast}=1.8$ is adequate, no investigation of
the high-velocity regime has been performed.
We assume that $\alpha_{ce}\lambda$ is in the range 0.1--1.0 (1.0
can be considered a very conservative estimate, giving the
minimum rate of LMXB formation through this process, whereas
0.1 may be a rather optimistic value), with 0.5 being our chosen 
standard value for calculations below.

\subsubsection{Tidal captures}

At periastron distances above
$R_{coll}$, and up to a few times $ R_{\ast}$, tidal capture can happen. 
\citet{Press} provided a way of calculating the energy absorbed
by the stars assuming $n=3/2$ polytropes and \citet{Lee} extended
these calculations to other polytropic indices. 
In this formulation, the energy of oscillations induced in the
non-degenerate star of mass $M_{\ast}$ during an encounter with a neutron star is
\begin{equation}
\label{eq:energy}
\Delta E_1=\left(
\frac{GM_{\ast}}{R_{\ast}}\right)^2\left(\frac{M_{ns}}{M_{\ast}}\right)^2\sum_{l=2,3}\left(\frac{R_{\ast}}{R_{p}}\right)^{2l+2}T_l(\eta )
\end{equation}
where $R_{p}$ is the distance of closest approach.
Only the spherical harmonic indices $l=2$ (quadrupole) and $l=3$ (octupole) 
are included, as higher indices give negligible contributions to
the energy \citep{Lee}. The parameter $\eta$ is defined as
\begin{equation}
\eta =\left(\frac{M_{\ast}}{M_{\ast}+M_{ns}}\right)^{1/2}\left(\frac{R_{p}}{R_{\ast}}\right)^{3/2}
\end{equation}
With the tabulated overlap integrals of \citet{Lee}, for polytropic
indices $n=3/2$ and $n=3$, we use the numerical method described
by \citet{Press} to calculate $T_l(\eta )$. From equation (\ref{eq:energy})
it is then possible to calculate the maximum value of $R_p$ (we call
this $R_{tid}$) for
which capture will occur, when the mass of the
star and the relative velocity at infinity is known. We use
this method for the case where the non-degenerate star is on
the MS. A polytropic index of $n=3/2$ is assumed for the 
mainly convective stars of mass, $M_*< 0.4~M_{\odot}$, and $n=3$
for stars $M_*> 0.4~M_{\odot}$ having radiative 
cores.\footnote{Note that this is different from the mass limit for
a fully convective star, $\sim 0.3M_{\odot}$ \citep[e.g.][]{Spruit},
used later as the mass at which magnetic braking ceases being effective. 
As properties of deep interiors of the star 
are not important for the tidal capture process,
a higher value is used in the latter case \citep[e.g.][]{Ray}}
For stars on the red giant branch, we use the
results of \citet{McMillan}, who calculated $R_{tid}$ for
captures of a neutron star by a 0.8 $M_{\odot}$ star at
various evolutionary stages along the RGB.
This mass is close to the MS turn-off masses for the
Galactic globular clusters (GCs), whereas the turn-off
mass in M31 is higher, $\sim$1 $M_{\odot}$. 
We use their results directly in both cases.
As tidal captures by evolved stars give a negligible 
contribution to the overall binary formation rates we did not attempt
to perform a more accurate computation for the case of M31.
We neglect tidal captures during the subsequent evolutionary 
stages. The structure of stars on the AGB is similar to those on
the RGB, but the time spent there is much shorter. The tidal capture
rate must therefore be lower. The time spent on the HB
is also very short, compared to the MS lifetime, and although the
tidal capture rate may be comparable to the RGB rate, it is
much smaller than the MS capture rates.

The tidal capture rate  $\gamma_{tidal}$ is computed as a rate of
encounter with the periastron distance  $R_{coll}<R_p<R_{tid}$,  
i.e. contribution of very close encounters resulting in collisions is 
subtracted:
\begin{equation}
\label{eq:tid}
\gamma_{tidal}=\int_0^{v_{max,t}}F(v_{rel},\sigma_v)[\sigma_{tidal}-\sigma_{coll}] 
v_{rel}dv_{rel}
\label{eq:gamma_tid}
\end{equation}
The upper integration limit $v_{max,t}$ is defined as the velocity
at which $R_{coll}=R_{tid}$ i.e. the term in square brackets is
required to be positive inside the integration limits. The
$\sigma_{tidal}$ is calculated from equation \ref{eq:sigma} with
$D=R_{tid}$.  

Important for the following evolution of tidal capture binaries
is the timescale on which the tidally induced oscillations are dissipated.
If this timescale is short \citep[as argued by][]{Kumar}, 
so that a large fraction of the energy is thermalized
within one orbital revolution of the binary, new oscillations will be
induced at each periastron passage. The binary quickly becomes
circularized with the final orbital separation (from conservation of
angular momentum): 
\begin{equation}
\label{eq:afinal}
a=\frac{(\sqrt{2G(M_{\ast}+M_{ns})R_p}+v_{rel}R_p)^2}{G(M_{\ast}+M_{ns})}
\end{equation}
For slow encounters (low $v_{rel}$ or small $R_p$), $a\simeq 2R_p$. 
If the dissipation time scale is too short, the quick conversion
of the energy (of the order of few per cent of the binding
energy of the star) may cause the star to expand and lead to a
merger. The outcome of the dissipation process depends on the
thermalization timescale and the region of the star where the energy
is deposited, both factors being unkown \citep{Podsiadlowski2}. 
Alternatively, if the dissipation  is
inefficient, coupling of the orbital motion with oscillation can cause  
large fluctuations in the orbital energy, substantially extending the
circularization process and potentially scattering a fraction of the
binaries to wider orbits \citep{Kochanek,Mardling}. 
The details and the final outcome of this processes  are poorly
understood. In the following we assume that all binaries become 
circularized with the final separation given by equation \ref{eq:afinal}. 

Equation \ref{eq:gamma_tid} defines the total tidal capture rate,
irrespective of the subsequent evolution of the tidally formed
binary. In order to calculate the rate of encounters, leading to 
formation of an LMXB, one needs to account for the finite braking
time scales. For this, $R_{tid}$, used to calculate $\sigma_{tidal}$ 
in equation   
\ref{eq:gamma_tid}, is replaced by min$(R_{tid},R_{brake})$, 
where $R_{brake}$ depends on the mass of the star and is chosen so
that the braking time scale for the tidal capture binary is shorter
than 5 Gyr. Due to small values of the tidal capture radius $R_{tid}$,
this does not affect the final rates significantly.

\subsection{Binary-single interactions}

The rate of exchange reactions between a binary $(M_1,M_2,a)$ and a
star $M_3$  is: 
\begin{eqnarray}
\gamma_{exch}(M_3,\sigma_v)=
\int f(M_1)  dM_1  \int p(q) dq
\nonumber
\\
\times
\int_{a_{min}}^{a_{max}} \gamma(M_1,M_2,M_3,a,\sigma_v)
\frac{dn}{da} da
\label{eq:gamma_exch}
\end{eqnarray}
where $f(M)$ is the distribution of mass of one of the stars in the
binary, $q$ is the binary mass ratio and $p(q)$ its probablity
distribution, $dn/da$ is the binary semimajor axis distribution and 
$\gamma(M_1,M_2,M_3,a,\sigma_{rel})$ is the exchange rate of 
a star $M_3$ into a binary $(M_1,M_2,a)$ computed from the equation
\ref{eq:rate1}. In the context of LMXB formation the third star $M_3$
is  an NS or a black hole.

\begin{table*}
\begin{tabular}{cccccccccccccc}
\hline
&&&&\multicolumn{5}{c}{without collisions}& \multicolumn{5}{c}{with collisions}\\
CO & $\sigma_v$ (1D) & $n_{\ast}$ & $f_{ns}$ & $N_{bin}$ & Total & MS-NS & RG-NS &
$\gamma_{ex}$ & $N_{bin}$ & Total & MS-NS & RG-NS & $\gamma_{ex}$\\
\hline
NS & 3 & $5\cdot10^{4}$ & 0.0025 & 1120790 & 3157 & 318 & 816 & 1.41$\cdot10^{32}$ & 2739700 & 3093 & 77 & 277 & 1.39$\cdot10^{31}$\\
NS & 15 & $3\cdot10^{5}$ & 0.0025  & 467000 & 1256 & 163 & 277 &  2.89$\cdot10^{31}$ & 1027400 & 1328 & 27 & 108
& 2.17$\cdot10^{30}$\\
NS & 150 & $10^{4}$ &  0.0068 & $1.68\cdot10^{8\dagger}$ & 1625 & 66 & 165 & 1.02$\cdot10^{29\dagger}$ &
$8.87\cdot10^{7\dagger}$ & 604 & 3 & 46 & 8.76$\cdot10^{27\dagger}$\\
BH & 150 &  $10^{4}$ & 0.0012 & $8.10\cdot10^{7\dagger}$ & 10938 & 82 & 739 & 1.49$\cdot10^{30}$ $^{\dagger}$ &
$8.10\cdot10^{7\dagger}$ & 9557 & 13 & 559 & 2.37$\cdot10^{29}$$^{\dagger}$\\
\hline
\end{tabular}\\
$^\dagger$ This simulation was performed for a limited range of orbital
separations and $\gamma_{ex}$ have been corrected for this.\\
\caption{Parameters and results for the three simulations of exchange
reactions.  The parameters are: the velocity dispersion $\sigma_v$ in
km s$^{-1}$, the number density of single stars $n_{\ast}$ in
pc$^{-3}$, the NS fraction in the population of  single stars $f_{ns}$
and the number of binaries simulated $N_{bin}$. The results  are:
the total number of NS binnaries formed through exchanges of a neutron
star in to a binary (Total) and the numbers of LMXBs formed in the 
simulations -- with MS (MS-NS) and RG (RG-NS) donors. The formation
rate $\gamma_{ex}$ calculated from equation \ref{eq:gamex} (cgs
units), is given for LMXBs with MS donors. The  other rates can be
computed  by scaling with  the numbers of formed
systems. The two groups of columns present results with and without
account for stellar collisions.}    
\label{tab:sim}
\end{table*}

A significant amount of effort has been invested in the past decades in 
studying binary-single interactions and  in calculating the encounter
crossections and rates.   
The three-body problem involved in encounters between a binary and
a single star can not be solved analytically and the computational
demand of the numerical solution has been prohibitive for the
studies of large ensembles of binaries based on direct
integration. 
An approach suggested and successfully implemented in 80-ies  was to
build large libraries  of interactions covering interesting range of 
initial parameters and, based on these libraries, to derive various
semi-analytical formulae describing  the interaction crossection and
outcome \citep[e.g.][]{Hut,Mikkola,Heggie3}.  This provided the
basis for computation of the elementary encounter rates
$\gamma(M_1,M_2,M_3,a,\sigma_{rel})$. 
Another ingredient required to compute the final encounter rates in an
ensemble of binary and single stars using eq.\ref{eq:gamma_exch} is
the semi-major axes distribution $dn/da$. This distribution has
complex time evolution, 
defined by the counterplay of binary (de-)excitation and ionization
processes which are difficult to take into account analytically, even
with the elementary crossections and rates given. 
Furthermore, unlike in collisions with red giants and in tidal
captures, the binary separation of a typical exchange binary is large
and one would have to take into consideration the subsequent evolution
of the binary parameters, before the Roche lobe contact is achieved
and an X-ray sources appears. 

Not surprisingly, the  Monte-Carlo methods has been proven to be more 
efficient. In these, the evolution of each individual binary is
followed through a number of encounters with single stars and, in some
implementations, with other binaries
\citep[e.g.][]{Hut,Heggie3,Sigurdsson,Davies1993}.
Due to computational limitations, these early simulations often relied
on libraries of interactions and semi-analytical crossections,
rather than direct integration of the three-body problem for each
interaction. 
While making possible to evolve sizable populations of binaries, this
approach has its deficiencies, as a number of distributions functions
(semi-major axes, eccentricities etc.) had to be replaced by average
values or treated in a simplified way \citep[e.g.][]{Hut}. 
Rapid advance in computing power and numerical methods  in recent
years have allowed full simulations, with each binary-single star
interaction being calculated exactly
\citep[e.g.][]{Fregeau,Ivanova4}. 
This lifts the assumptions and approximations
mentioned above, and it is this method that we employ in the current 
study. 
This approach has been implemented as early as in 90-ies
\citep[e.g.][]{Portegies1997}, but the moderate numbers of binaries
achievable then have been insufficient to 
study the formation of LMXBs (see table \ref{tab:sim}).
The next level of complication (and of computational demand) is a complete
time dependent simulation of an N-body system composed of binary and
single stars with realistic mass functions. 
Presently this is becoming feasible, but is still
limited to systems containing $\lesssim$ 100.000 stars
\citep{Portegies2007,Hurley2007}.

\subsubsection{Monte-Carlo simulations}


In our simulations we follow the evolution of binaries
in an environment of single stars, with special emphasis on
interactions between the single stars and the binaries.
The outcome gives a reasonable indication of the importance of
this process, compared to the two paths of dynamical formation
of LMXBs from single-single encounters discussed above.
The simulations are based on the FEWBODY code of
\citet{Fregeau}. FEWBODY numerically 
integrates the orbits of the stars during the interaction, and
automatically classifies and terminates calculations as soon as the
outcome is unambiguous, which makes it well suited for carrying
out large sets of binary interactions.

All binaries and single stars are assumed to be formed at the
same time, and the simulation of the binaries begins 0.5 Gyr 
after the star formation episode. The masses of single stars are
assumed to follow the initial mass function of \citet{Kroupa}. 
Only main sequence stars and neutron stars are included, and the
main sequence turn-off mass evolves with the age of the population
\citep[as given by][]{Girardi}.  
The number density of single stars is kept constant during the
simulation. 

The initial binaries are drawn randomly from a population with
properties typical of binary population synthesis studies
\citep[e.g.][]{Dewey,Pols}. The
mass distribution of the primary stars ($M_p$) was chosen 
to be the same as the mass function of single stars, while the mass of
the secondary chosen from a flat mass ratio distribution.
The distribution of orbital separations $a$ was assumed flat in
$\log a$ between a minimum separation corresponding to one of the
stars filling its Roche-lobe, and a maximum separation of $10^4 R_{\odot}$,
consistent with the distribution found by \citet{Duquennoy} in this range.
The initial eccentricity of the binaries was set to 0.
Each binary is evolved for 15 Gyr in the single star environment,
taking into account stellar evolution and evolution of the binary
orbit due to magnetic braking and gravitational radiation
as well as encounters with single stars \citep[but omitting the more complicated late phases of binary
evolution of more advanced models, e.g.][]{Dewey,Pols,Portegies1996,Hurley,Belczynski} .

The binaries are evolved in timesteps of maximally 0.01 times the
average time between encounters with single stars, where an
encounter is assumed to happen if a star comes closer than 6
orbital separations. For each timestep binary parameters are
adjusted according to gravitational wave emission \citep{Landau,Peters}
and magnetic braking \citep{Rappaport}. We assumed the
disrupted magnetic braking model, where magnetic braking is
ineffective when the MS star is totally convective \citep{Rappaport,
Spruit}. This is the case when the mass of the star is below 
$\sim$ 0.3 $M_{\odot}$.
The probability of an encounter between the binary and a single
star within a timestep of length $\Delta t$ is given by a weighted
average over the distribution of relative velocities and over
the mass function of single stars.
\begin{equation}
P_{enc}=  \Delta t \ n_{\ast}\int_M\gamma f(M)dM 
\end{equation}
Here $n_{\ast}$ is the number density of single stars, $\gamma$ is
given by equation \ref{eq:rate1}, with the cross-section found from
equation \ref{eq:sigma} with $D=6a$, where $a$ is the orbital separation
of the binary.
Random numbers are drawn to
see whether an encounter occurs. If this is the case, the parameters
of the encounter are drawn from their respective
probability distributions. The mass of the single star is
drawn from the mass function (allowing for neutron stars
also). The probability of an encounter distance $D$
is proportional to $\frac{d\sigma}{dD}$ (where $\sigma$ is given by 
equation \ref{eq:sigma})
out to the maximum
distance of 6 orbital separations of the binary.
The distribution of encounter
velocities $v_{rel}$ is given by equation \ref{eq:veldis}. 
The encounter is then solved
for using the {\tt binsingle} program of FEWBODY. Binary phase and
encounter angle is chosen randomly by FEWBODY from a flat and
an isotropic distribution, respectively.
The simulation of a binary is terminated when 
one of the following occurs: (1) the binary is disrupted, (2) one
of the stars evolves off the main sequence or (3)  Roche-lobe
contact is reached. If one of the binary components is an NS,
possibility 3 leads to the formation of an LMXB. Possibility
2 also leads to Roche-lobe overflow, but as discussed above,
such RGB-NS systems are shortlived and transient X-ray sources.

The simulations are performed with several simplifying assumptions.
We discuss the most important of them below. 
Encounters between binaries
are ignored. As most wide binaries are quickly destroyed, the
binary fraction decreases fast and binary-binary encounters should
only matter at early times. Moreover, in most binary-binary
encounters, two or more of the stars merge \citep{Fregeau}.
Secondly we have neglected the effect of tidal 
interaction in the evolution of the binaries \citep{Zahn1,Zahn2}. 
This will tend to lock the rotation of the stars to the orbit and to
circularize the orbit, thus decrease somewhat the time it  
takes for a system to achieve Roche-lobe contact. The significance of
this effect is difficult to estimate, for its implementation in 
population synthesis codes, see \citet{Belczynski}.
Evolved stars were not included in the simulations.
For the single star population this should not be a problem,
as an encounter between an evolved star and a binary will
probably lead to a merger of some sort due to the
large radius of the evolved star. The net effect of such encounters
will most likely be a decreased binary fraction. 
As for the evolved stars in the binaries, they will 
lead to Roche-lobe overflow. It is unlikely that a neutron
star can be exchanged into such a system without the
occurrence of a physical collision. We verified with
test simulations that close encounters between tight binaries and 
single stars in which one of the stars is evolved in almost all
cases lead to merger of two or all three stars, 
in accordance with the conclusions of \citet{Fregeau}.

\subsubsection{Results of simulations}

We performed three simulations with different velocity dispersions
and densities, to cover  the environment in both M31 and in GCs. 
Parameters and results of the simulations are summarized in table
\ref{tab:sim}. 
Presented in the table are the numbers of neutron star binaries 
created in the simulations --  the total number and
the numbers of Roche-lobe filling systems. The latter is divided into
the following two categories: 
the binaries in which Roche-lobe overflow occurs due to evolution of
the binary orbit, while the companion star is on the MS (MS-NS) and 
the  systems in which the mass transfer is initiated due to the
evolution of the companion star off the main sequence (RG-NS).

\begin{figure}
\includegraphics[angle=270]{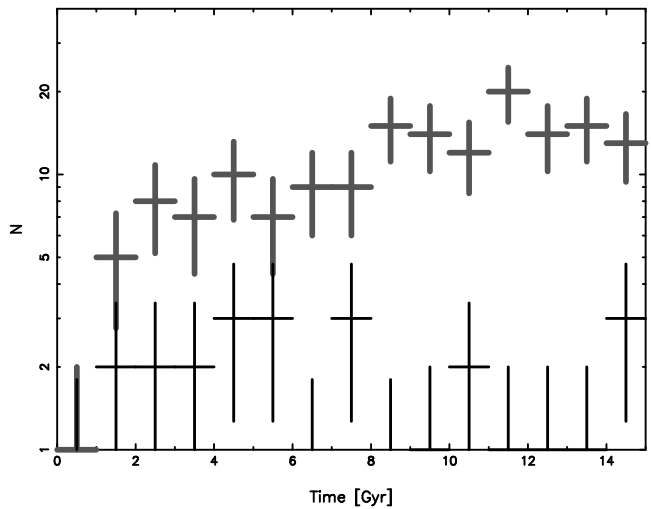}
\caption{The evolution of the dynamical formation rate of LMXBs due to
exchange reactions in a low velocity environment of a globular cluster
($\sigma_v=15$ km s$^{-1}$).  
The plot shows the number of systems in which Roche-lobe contact was
reached, per time bin, as a function of time from the start of the
simulation.  Results of simulations without  and with
account for mergers are shown  with thick grey and thin black crosses
respectively. 
}
\label{fig:times}
\end{figure}

We convert the numbers into $\gamma_{ex}$ rates, which can be
directly compared to the single-single interaction rates computed
in the previous section, using
\begin{equation}
\label{eq:gamex}
\gamma_{ex}=\frac{\Gamma}{n_{bin}n_{ns}}=\frac{N}{T_{sim}n_{ns}N_{bin}}
\end{equation} 
where $n_{bin}$ and $n_{NS}$ are the number densities of binaries and
neutron stars, $N_{bin}$ is the total number of binaries in the
simulation and $T_{sim}$ is the simulations time span. 
Note, that with the above definition, $N_{bin}$ is the total number of
binaries simulated, i.e. $n_{bin}$ has the meaning of the primordial
volume density of binaries.
The rates for MS-NS systems are  given in the Table
\ref{tab:sim}. The other rates can be computed from these by scaling
according to numbers of binaries. The Monte-Carlo uncertainties can be
estimated assuming a Poissonian distribution for the numbers of
binaries.
In the M31 simulations (the largest value of $\sigma_v$)
we simulated binaries in a limited range of separations, as wide
binaries are quickly ionized and do not contribute to the LMXB
production rates. The final value of the exchange rates given in
Table \ref{tab:sim} has been corrected correspondingly.

In the initial simulations the radii of stars were set to zero,
i.e. the possibility of stellar collisions was not accounted for.
We performed a second set of simulations, in which two stars
with radii $R_1$ and $R_2$ were assumed to collide if
the distance of closest approach was $D<1.8(R_1+R_2)$. 
We assumed that collisions lead to a merger and
removed from simulations all binaries that experience such
events. With these assumptions, the rates of LMXB formation 
decrease dramatically, by about an order of magnitude, i.e.
in $\sim 90\%$ of binary-single interactions
which could potentially lead to an  exchange of the neutron star into
the binary, two or more of the stars collide. This result is in
agreement with \citet{Fregeau} and demonstrates that 
mergers are a determining  factor in  the formation of exchange LMXBs.  

Figure \ref{fig:times} illustrates the time dependence of the
formation rate of exchange LMXBs in a low velocity environment of a
typical globular cluster (simulation with
$\sigma_v=15$ km s$^{-1}$). Shown in the figure is the number of
binaries per time bin, in which the Roche-lobe overflow was initiated
during the given time bin, irrespective of the time when the neutron
star was exchanged into the binary.
There is an obvious increase with time,
due to the fact that in a  low velocity dispersion environment most
MS-NS binaries 
are created with relatively large 
orbital separations and need to be hardened by further collisions in order
to become LMXBs. This is in contrast to M31 (large velocity
dispersion), where the rate is constant, due to the fact that almost
all  exchange  LMXBs there are formed from binaries with small orbital
separations,  $a \lesssim 10 R_{\odot}$. For such binaries,
binary-single interactions are not an important factor in their
further evolution towards Roche-lobe overflow. Also shown in the plot
by thin crosses is the result of simulations with account for mergers.

For interactions, where the final binary is harder than the
initial binary, the binding energy lost is converted to kinetic
energy of the binary and the single star. The velocity of the
binary obtained due to this effect is often referred to as the
dynamical recoil velocity\citep[e.g.][]{Sigurdsson,Davies1993}. 
We find that in globular clusters,
the binaries that end up as LMXBs typically undergo several
encounters, in which recoil velocities in the range 
$\sim 30-50$ km s$^{-1}$ are obtained. The effect of this is
that the binary is ejected from the core (and sometimes also
from the GC) to the less dense regions of the GC, where dynamical
interactions are rare. After a significant time, the binaries will
re-enter the core due to mass segregation. For a recent discussion
of this binary cycling in and out of the core of a GC using N-body
simulations, see \citet{Hurley2007}.
As this cycling can lead to significant prolongation of the binary
lifetime before the formation of a LMXB, it decreases the formation
rate due to this channel. In the simulations with
physical collisions this effect is smaller, due to the fact that
many of the encounters that lead to high recoil velocities are also
the encounters that lead to collisions. We note that the encounter
cross-sections of LMXBs are so small that only a small fraction
of them experience significant encounters, and they are therefore
retained in the GC cores. 
In M31 most LMXBs are formed through only one encounter, an exchange
reaction in which the orbital separation is decreased significantly,
for an already tight binary. The recoil velocities are therefore
extremely high, typically 100-600 km s$^{-1}$. Even for the deep
potential wells of galactic bulges, such velocities can be enough
to eject the binaries. However, as will be shown in section
\ref{sect:realistic} this channel is not a significant source of
LMXBs in M31 anyway.

\subsection{Comparison of the rates}
\label{sect:comparison}

\begin{figure}
\resizebox{\hsize}{!}{\includegraphics[angle=270]{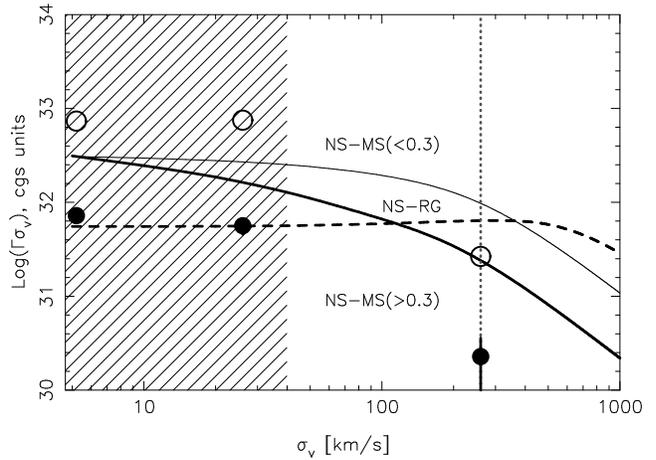}}
\caption{Comparison of NS LMXB dynamical formation rates as a function
of the stellar velocity dispersion.
Shown are the rates $\gamma\sigma_v$ for tidal captures by main
sequence stars $>0.3M_{\odot}$ (thick solid line) and $<0.3M_{\odot}$
(thin solid line), collisions with evolved stars (assuming $\lambda\eta=0.5$, 
dashed line)
and exchange reactions (circles with error bars). For the exchange
reactions, results are shown for simulations without 
(open symbols) and  with  (filled symbols) account for collisions. Error
bars are only shown if they are larger than the circles. The
hatched area shows the velocity dispersion range typical for globular
clusters, the dotted line for the inner bulge of M31.} 
\label{rates1}
\end{figure}

The results of this section are summarized in figure \ref{rates1}
where  we compare the rates for the three main LMXB formation processes,
involving neutron stars, as a function of the stellar velocity dispersion.   
In computing the rates for the tidal capture and collisions with evolved
stars we assumed an environment (IMF, age metallicity etc.) similar to
the bulge of M31, as described in section \ref{sect:realistic}.
As will be discussed in section \ref{sect:realistic} the stellar
environment in GCs is significantly different from that in M31, 
in particular with regard to the present day mass function and metallicity,
and this is taken into account in our final estimates.
The goal of this section is to highlight the influence of
the velocity dispersion.
The exchange LMXB rates are from the simulations of the previous
subsection, without and with account for mergers. 
It should be noted, that low and high velocity parts of these
simulations were tailored for GC and M31 environment
respectively, therefore were performed for different values of the
stellar density, main sequence cut-off mass $M_{co}$, age and
metallicity. For this reason,  
although they do correctly illustrate the general trend of 
the exchange rates with the stellar velocity dispersion, they should not
be used to study the exact dependence. One should also keep in mind,
that in order to convert the formation rates into the numbers of
X-ray sources, the LMXB life-time considerations should be taken into
account, as discussed in the section \ref{sect:lifetimes}. 

Figure \ref{rates1} illustrates significant velocity dependence
of the relative importance of different LMXB formation channels and
suggests that the relative contributions of different subclasses of LMXBs
should be different in GCs 
and in the galactic centres. In the low velocity  environment of a
GCs all three 
processes make comparable contributions to the LMXB production rates
(but not to the numbers of X-ray sources observed in any given time, see
below), with some prevalence of tidal captures by the main sequence
stars, depending on the exact value of the velocity dispersion.    
In the high velocity environment of a galactic bulge, on the contrary,
the tidal capture by main sequence stars with $M_*>0.3M_{\odot}$ and
exchange reactions are unimportant and the LMXB formation rates are
dominated by the collisions with evolved stars and tidal captures by
very low mass stars.
However, the comparison between globular clusters and M31 is
more complex than comparison of the velocity dependent rates, as these
environments also differ in other properties of the stellar
populations, such as the present day mass function, metallicity,
binary fraction etc. 
This is considered in detail in section \ref{sect:realistic}.

Finally, the total rate of encounters in volume $V$ can be obtained
as: 
\begin{equation}
R=\int_V\left(\frac{\rho_{\ast}}{<M>}\right)^2 f\int \gamma
(M,M_{ns},\sigma_v)f(M)dMdV
\label{eq:rate}
\end{equation}
where $f=f_{ns}, f_{bh}, f_{bin}$. Note, that the former two
coefficients refer to the present day values, while the latter is the
primordial binary fraction, as clarified in the previous subsection.

\subsection{Black hole encounters}
\label{sect:bh}

The discussion above was limited to the formation of LMXBs in
which the compact object is an NS. Of course, the same
processes are relevant for black holes (BHs), and these are considered 
below.  
We assume that stars with initial masses in the $30-100 M_{\odot}$ range 
become BHs, with a canonical mass of 10 $M_{\odot}$. The rates of
tidal captures and and collisions with evolved stars can then be found
from the equations of the previous sections, 
replacing $M_{ns}$ with $M_{bh}$ and $f_{ns}$ with $f_{bh}$.
With the initial mass function of \citet{Kroupa}\footnote{
To use the mass function of \citet{Kroupa} to estimate $f_{bh}/f_{ns}$
is appropriate for the bulge of M31, where mass segregation is
negligible, except for the inner few parsec \citep{Freitag}. In
GCs this is not the case as three factors affect the mass function, 
namely the mass segregation causing more massive
objects to sink to the core, supernova kicks depleting the GCs of
NSs and the ejection 
of BHs due to encounters with other BHs \citep{Portegies2002}.
This is discussed in detail in section \ref{sect:realistic}
}, 
$f_{bh}=0.17\cdot f_{ns}$.
Note that although there are $\approx 6$ times fewer BHs than NSs,
this is countered by the  the gravitational focusing term
(equation \ref{eq:sigma}) which makes the encounter cross-section
$\sim$5 times larger for a BH. 
As for the neutron stars, we assume that  $R_{coll}/R_{\ast}=1.8$.
We note, however, that from equation \ref{eq:afill} it might be
expected that $R_{coll}$ may be  larger for the very small mass ratios
considered here.

\begin{figure}
\includegraphics[angle=270]{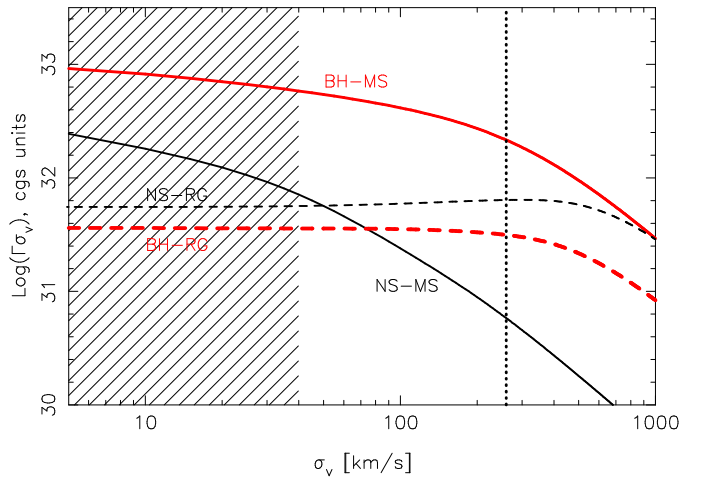}
\caption{Comparison of the LMXB formation rates in encounters with
neutron stars (thin lines) and black holes (thick lines). 
Shown are tidal captures by main sequence stars of mass
$>0.3M_{\odot}$ (solid lines) and collisons with RGB/AGB stars (dashed
lines). The normalization of the BH rates have been multiplied with
$f_{bh}/f_{ns}$. 
 }
\label{fig:BH}
\end{figure}

When considering various aspects of single-single encounters on the
basis of the total energy arguments it should be noted that the
kinetic energy at infinity in the centre-of-masses frame is defined by
the reduced mass $\mu=M_1 M_2/(M_1+M_2)$ which, for the low mass stars,
only depends weakly on the compact object mass. Therefore, even if the
energy equipartition is inefficient and the NS and BH velocity
dispersions are comparable, as is the case in the M31 bulge,
the kinetic energy at infinity, $\mu v_{rel}^2/2$, will not be much
higher in the case of a black hole.
For this reason  the energy considerations in collisions with evolved
stars (section \ref{sect:coll}, eq.\ref{eq:CE}),  will not
change significantly and the fraction of collisions expected to lead
to a bound systems for a given velocity  dispersion depends only weakly
on the mass of the compact object.

For tidal captures, the energy of the oscillations
induced in the non-degenerate star is roughly $\propto M_{NS,BH}^2$
(see equation \ref{eq:energy}), and the capture distance is therefore
larger for BHs than for NSs. At low velocities, this can enhance the
rate by a moderate factor of $\lesssim$2.  At large velocities, on the
contrary,  $R_{tid}\sim R_{coll}$, and even a small increase in
$R_{tid}$ can  drastically increase  $\sigma_{tid}-\sigma_{coll}$ and
thereby the overall tidal capture rate. 
As the total energy  to be
dissipated in the interaction depends weakly on the compact object
mass, the impact of the tidal oscillation on the thermal state of the
normal star does not become more severe for the black hole. Therefore,
if a tidal capture is possible  at all, it is possible for black holes
as well as for neutron stars.

There is one important difference
between BHs and NSs,  namely that the retention factor for black holes
in globular clusters is close to zero \citep{Portegies2002}. 
For this reason, black holes do not contribute to
dynamical LMXB formation in globular clusters. In M31, however, the
black hole fraction should be close to 
the IMF-based estimate given above, due to much longer energy
equipartition time scale than in globular clusters.

In figure \ref{fig:BH} we compare the LMXB formation
rates in single-single encounters involving black holes and neutron
stars. Obviously, black holes can make sizable contribution to the
LMXB formation rates, especially in the high velocity regime.

\subsection{Numbers of X-ray sources}
\label{sect:lifetimes}

In order to convert the encounter rates to the numbers of X-ray
sources observed at any given moment of time, one needs to consider
the evolution of a binary through the X-ray phase. A definitive answer
can be obtained from proper population synthesis calculations, which
is beyond the scope of this paper. In a simpler approach one may
consider characteristic life times $\tau_X$ of binaries at different
phases of its evolution. The number of X-ray sources $N_X$ can be then
related to the corresponding encounter rate: 
$N_X\sim \gamma\,\tau_X$. 
Taking into account dependence of the $\tau_X$ on the mass and
evolutionary status of the donor star and their mass distribution,
we obtain an expression, similar to
the equation \ref{eq:rate} for overall 
encounter rate:
\begin{equation}
N_X\approx 
\int_V\left(\frac{\rho_{\ast}}{<M>}\right)^2 f\int \tau_X(M)\,
\gamma (M,M_{ns},\sigma_v)f(M)\,dMdV
\label{eq:nx}
\end{equation}
where $f$ is defined as in eq.(\ref{eq:rate}) and 
$\tau_X(M)=\Delta M_d/\dot{M}$, $\Delta M_d=M_i-M_f$, $M_i$ 
is the initial mass of the donor star and  $M_f$ its final mass
in the given evolution stage (e.g. for a star with
initial mass $>0.3 M_\odot$, $M_f=0.3 M_\odot$ -- the mass
corresponding to the period gap).
In case of an LMXB powered through the Roche-lobe overflow, the
$\dot{M}$ is defined by the orbital braking
mechanism and the mass-radius relation for the donor star. 
The stability of the mass transfer in the accretion disc should be
also taken into account in order to identify persistent/transient
nature of the binary. The integral in eq.(\ref{eq:nx}) is taken over
the range of the masses relevant to the given type of X-ray binaries. 

\begin{figure}
\includegraphics[angle=0]{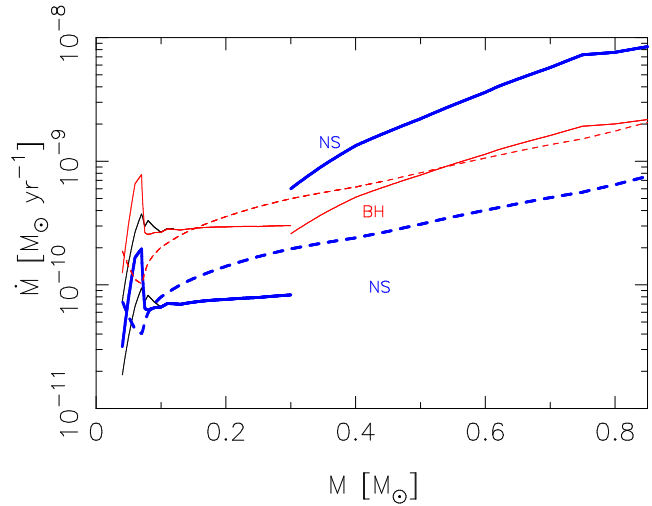}
\caption{Dependence of the mass accretion rate $\dot{M}$ in a
Roche-lobe filling system  on the mass of the donor star. 
The curves for a neutron star and a black hole binary are shown by
thick and thin lines. The calcultaions based on 10 and 1 Gyr
isochrones give identical result down to $\sim 0.1 M_\odot$, below
which the 10 Gyr isochrones give more pronounced spike in $\dot{M}$.  
The transiency limits are plotted by respective dashed lines.
The method of calculations and assumptions are
described in section \ref{sect:lifetimes}
 }
\label{fig:mdot}
\end{figure}

Below we examine evolution and characteristic values of $\dot{M}$ of
X-ray binaries formed formed via different dynamical processes
considered in in this paper.  We accept the standard prescriptions for
the magnetic braking \citep{Rappaport} and
gravitational radiation \citep{Landau,Peters} and the transiency 
criterium in the form
published by \citet{Dubus} for irradiated discs. One should keep
in mind that these simple presciptions predict time averaged
quantities but may fail to explain the momentary values of
luminosity, which may vary significantly on the timescales of
days--months--years. The dependences of the mass accretion rate on the
mass of the donor star for NS and BH binaries are shown in
figure \ref{fig:mdot}. These dependences were computed based on standard
formulae for a Roche lobe filling secondary \citep{Heuvel} assuming 
that the secondary is in the thermal
equilibrium. As was demonstrated by \citet{Stehle}, this assumption 
gives sufficiently accurate results for the
main sequence donor. For the mass-radius relation we used the 10 Gyr
isochrones of \citet{Baraffe1} and \citet{Baraffe2} for stars $M<0.1M_\odot$
and  $M>0.1M_\odot$, respectively. Also shown in figure \ref{fig:mdot} are 
transiency limits for different types of compact object. The NS and BH
masses were assumed $1.4$ and $10$ $M_\odot$ respectively.
The spike in $\dot{M}$ at $\approx 0.07M_\odot$  is caused by the
steepening of the mass-radius relation just above the hydrogen burning minimum
mass, as given by the isochrones. This is due to the fact that
correlation effects between particles becomes important, and the deviations
from an ideal gas decreases the pressure (I. Baraffe, private communication). 
Below the hydrogen burning minimum
mass, degeneracy effects dominate and the mass radius relationship
becomes $R\sim M^{-1/3}$.
The spike is less pronounced in the 
1 Gyr isochrones (shown in figure \ref{fig:mdot} by thin solid lines) which
might be more appropriate for the thermal state of a mass-losing brown
dwarf.

{\em Collisions with evolved stars.}
In a collision with a red giant, an
ultra-compact X-ray binary (UCXB) with a He white dwarf donor is 
formed. The white dwarf mass equals approximately the mass
of the red giant's core \citep{Lombardi}, i.e. is in the 0.1-0.4
$M_{\odot}$ mass range, depending on the evolutionary stage of the red
giant.  
The evolution of such a system  includes a very fast 
initial stage of very short, $\sim$ minutes, orbital period and very
high, super-Eddington $\dot{M}$. During this period the white dwarf
donor is quickly reduced  to a $\sim 0.06 M_\odot$ after
which a more ``normal'' UCXB with $P_{orb}\sim 10$ min and 
$L_X\la 10^{38}$ erg s$^{-1}$ emerges, similar to the ones  observed in our
Galaxy.     
Overall, such a system will spend $\sim 0.1$ Gyr with the luminosity
$10^{36}-10^{38}$ erg s$^{-1}$, before the white dwarf is depleted below
$\approx 0.02 M_\odot$. Somewhere around this mass the sources will
cross the stability threshold and will become transient
\citep{Deloye,Bildsten}.

The cores of less evolved, sub-giant stars are not fully degenerate
and/or hydrogen-depleted. In this case a collision
will result in a binary with He or brown dwarf donor, depending on the
core mass and chemical composition. Such a binary is also driven by
gravitational radiation, 
but due to the larger radius the period minimum is higher, $\sim 20-30$
minutes, and super-Eddington mass transfer does therefore not occur. 
For such systems, a life time of
$\sim 200-300$ Myrs may be expected (N.Ivanova, private communication). 
In order to make a crude estimate of their fraction  we 
assume that the core of an RGB star becomes fully degenerate, when the
central density exceeds $\rho_c\ga (5-10)\,\rho_{crit}$, 
where $\rho_{crit}$ is the critical density above which electron gas is
degenerate  ($\rho_{crit}\sim 2.4\cdot 10^{-8}\mu_E T^{3/2}$ g cm$^{-3}$).
We estimated from the Padova stellar tracks that this occures at
stellar radii of $R\sim (3-5)\times R_\odot M/M_\odot$. As discussed
in section \ref{sect:coll}, given the high stellar velocities in M31,
only RGBs with rather  small radii can effectively capture a compact
object through collisions, and we expect that in a large fraction,  
$\sim 50-80$ per cent, of X-ray sources created through this mechanism
the donor star is not fully degenerate. 
In the low velocity environment of globular clusters this
fraction is smaller, $\sim 25-40$ per cent.

{\em Tidal captures by main sequence stars with $M>0.3M_\odot$} lead
to formations of ``usual'' LMXBs, similar to the ones constituting the
majority of systems with main sequence donors observed in the Galaxy. 
These sources are driven by the magnetic braking and luminosities of
$\sim 10^{36.5-38.0}$ erg/sec and lifetimes of $\sim 0.1-0.5$ Gyr should
be expected \citep[e.g.][]{Heuvel}. Note, that these estimates depend critically on
the magnetic braking prescription, the weak magneting breaking
predicting up to several times smaller luminosities and longer life
times \citep{Ivanova3}.  From figure \ref{fig:mdot} it can be seen that 
all black hole systems are expected to be
transient,  in agreement with BH binaries statistics in the Milky
Way.

{\em Tidal captures by  main sequence stars of very low mass,
$M<0.3M_\odot$.} 
For these fully convective very low mass stars the magnetic braking
is believed to be inefficient \citep{Spruit}, therefore the accretion 
is driven by
the gravitational radiation. From the requirement that the donor
star fills its Roche-lobe we have that the orbital periods of these
systems are in the $\sim$hours range, and that gravitational radiation 
can provide luminosities of $\sim 10^{36.0-36.5}$ and  $\sim
10^{36.5-37.0}$ erg/s for NS and BH systems respectively
\citep[see also e.g.][]{Podsiadlowski,Yungelson}.
>From figure \ref{fig:mdot} it can be seen that the systems with 
$M\ga 0.15M_\odot$ will be transient, these 
constraints being more severe for the NS binaries.
Integration of the mass transfer rate gives that the life times during 
the persistent phase are $\sim 300$ Myr, and that 
the life times during the transient phase are $\sim 1$ and $\sim 4$
Gyrs for BH and NS systems respectively.

It is interesting to consider the final stage of evolution of these
systems, after the donor star is reduced to $\la 0.1M_\odot$, below
the nuclear burning limit. As  these are descendants of very low
mass stars, whose nuclear time scale is much longer than the
cosmological time, they consist mainly of hydrogen and they will 
become brown dwarfs.
Given the mass-radius relation for brown dwarfs, the mass transfer
rate drops quickly when the mass reaches 0.05 $M_{\odot}$
(Fig.\ref{fig:mdot}), and these systems
become transients, 
similar to some of the accreting msec pulsar systems, observed in our
Galaxy.

We note that in the binary systems with very low mass ratios,
q$\lesssim0.02$, the circularization radius exceeds the tidal
truncation radius \citep[e.g.][]{Paczynski}.
It is therefore not entirely clear whether the stable mass transfer is
possible, see e.g. discussion in \citet{Yungelson} (section 3.3).
Such low mass ratios can be reached for the most low mass black
hole systems.

\section{M31 and the Milky Way globular clusters}
\label{sect:realistic}

Below we compute rates of dynamical formation of LMXBs and their
expected numbers in M31 and in Galactic GCs. 
For this, we need to specify velocity dispersion, initial and present
day mass functions, age, metallicity and stellar isochrones. 
These parameters are different in GCs
and galactic centres. The difference in stellar
velocities is an important one, as discussed in section
\ref{sect:comparison}, but several other properties of
stellar populations play equally significant roles in shaping the
population of dynamically formed binaries. The factor of prime
importance is highly efficient mass segregation in GCs.
Its two most significant consequences are: 
\begin{enumerate}
\item The present day mass function. Due to efficient mass 
segregation, the inner regions of the GCs, where most of
the encounters happen, are depleted of low mass
stars, to the degree that the mass function is essentially flat
\citep[e.g][]{King3,Marchi,Albrow}. 
This is not the case for a galactic bulge, where the mass
distribution of main sequence stars is sufficiently
well represented by the Kroupa IMF \citep{Zoccali}. As a result,
the tidal captures by very low mass stars, dominating the binary
formation processes in M31 (Fig.\ref{rates1}),  are significantly less
important in GCs.   
\item Abundance of BHs. 
GCs are believed to be depleted of black holes
\citep{Portegies2002}, due to  mass segregation and BH-BH encounters
(although the observation of an ultra-luminous X-ray source in a
GC in NGC 4472 by \citet{Maccarone2}, indicates that some BHs may be
present in GCs). 
Therefore
tidal captures of BHs do not play any roles 
in globular clusters as opposite to the case of M31. Note that in the
latter case the role of black holes is further
enhanced by the velocity dependence of the tidal capture
cross-section, as discussed in section \ref{sect:bh} and shown in
figure \ref{fig:BH}. 
\end{enumerate}

Among other factors, leading to further quantitative differences,
the following should be mentioned: 
(i) Due to supernova kicks \citep{Lyne}, large fraction of neutron stars escape the
parent cluster, with the NS retention factor being in the $\sim
0.1-0.2$ range \citep{Drukier,Pfahl}. On the other hand, the mass segregation
of the remaining NSs may increase their density near the
globular cluster centres, thus compensating for the low retention
fraction. 
(ii) Binary fractions are different in globular clusters and
galactic centres, due to different rates of binary-single processes, 
caused by difference in velocities and stellar densities. This 
is important for exchange rates and is taken into account in our 
simulations automatically.
(iii) Finally, different ages and, especially metallicities result in
different mass-to-light ratios, the main sequence turn-off mass and
duration of the red giant phase, as discussed in section
\ref{sect:gc}. 

For these reasons the comparison between globular clusters and M31 is
more complex than comparison of the velocity dependent rates, shown in
figure \ref{rates1}. It is the subject of this section.

\begin{figure}
\begin{center}
\resizebox{\hsize}{!}{\includegraphics[angle=270]{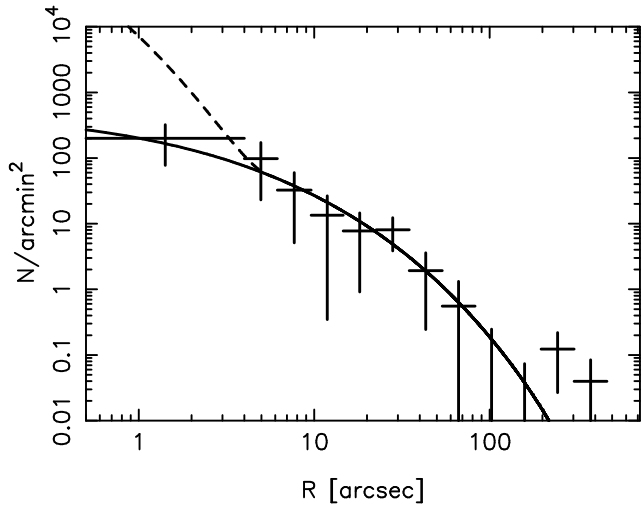}}
\caption{The radial distribution of ``surplus'' X-ray sources
computed as a difference between the data and best fit model in
figure \ref{fig:fit}.
The solid line shows the projected  $\rho_*^2$
distribution, computed from the original mass model of the M31 bulge
from \citet{Riffeser}.
The dashed line was computed from the mass distribution with the
circumnuclear stellar disc excluded.
Both model distributions are normalized to the observed number of
surplus sources outside 4 arcsec.} 
\label{fig:surplus}
\end{center}
\end{figure}

\subsection{M31}
For the stellar models we used an ischrone with a metallicity $\sim$
1.5 times solar \citep{Sarajedini} and an age of 12 Gyr \citep{Rich}\footnote{Isochrone file
isoc\_z030.dat from \citet{Girardi}.}.  This
gives a main sequence turn-off mass of 0.9532 $M_{\odot}$ and a 
mass at the tip of the AGB of 1.0081 $M_{\odot}$ 
(initial masses).\footnote{We note that the precision of these masses 
is given for identification on the published stellar tracks. This should
not be confused with the precision to which these values are actually
know, which is much lower.}
Stars more massive than this have all turned into stellar remnants.

\begin{table*}
\begin{center}
\begin{tabular}{llcccccc}
\hline
Object & Type & MS($<0.3 M_{\odot}$) & MS($>0.3 M_{\odot}$) & RGB & HB & AGB &
Exchange\\
(1) & (2) & (3) & (4) & (5) & (6) & (7) & (8)\\
\hline
NS & Tidal capture & {\bf 15.1} (15.7)& {\bf 0.8} (0.9) & 0.01 & - & - & 3.5\\
NS & Collisions & 36.6 & 46.6 & {\bf 6.5-13.3} & 5.2 & {\bf 0.01-0.67}
 & {\bf 0.3}\\
\hline
BH & Tidal capture & {\bf 65.3} (65.3) & {\bf 14.2} (14.2) & 0.09 & - & - & 8.8\\
BH & Collisions & 37.7 & 38.0 & {\bf 3.1-5.8} & 2.0 & {\bf 0.00-0.06}& {\bf 1.0}\\
\hline
\end{tabular}
\caption{Formation rates of LMXBs in M31, per Gyr.
The entries in bold are expected to lead to the formation of LMXBs.
The columns are:
(1) The type of compact object; 
(2) the capture process;
(3) rate of interactions with MS stars of mass $<0.3 M_{\odot}$, only those
initiating Roche-lobe overflow within 5 Gyr are included (full number
is given in the parenthesis);
(4) rate of interactions with MS stars of mass $>0.3 M_{\odot}$, same criteria
as (3);
(5) rate of interactions with stars on the RGB, for collisions only
those with $a_f<5R_{\odot}$ are included for $\eta\lambda=0.1-1.0$;
(6) rate of interactions with stars on the HB, tidal captures were not
calculated;
(7) rate of interactions with stars on the AGB, tidal captures were not
calculated, same criteria as (5);
(8) LMXBs created through exchange reactions, estimated from simulations
without collisions (in tidal capture rows) and with collisions (in collision
rows).
\label{tab:M31rates} }
\end{center}

\end{table*}

The velocity
dispersion (3D) was assumed to be constant, $\sigma_v$=260 km s$^{-1}$
\citep{McElroy,Widrow}.
The density structure of M31 was constructed using the model of
\citet{Riffeser}, based on the Gunn-r band photometry presented by
\citet{Kent}. In this model the total \textit{R}-band luminosity of the
bulge out to a distance of 12 arcmin from the centre of M31 is
1.18 $\cdot 10^{10} L_{\odot,R}$.
We normalized the density by requiring the integrated 
R-band luminosity over the mass function (giving a mass-to-light ratio
of $(M/L)_R$=3.27) to match the R-band luminosity
of the model, giving the bulge a total mass of 3.9$\cdot 10^{10}
M_{\odot}$ of stars in the 0.1-1.0081 $M_{\odot}$ range. 
The projection of this model agrees with the mass
distribution inferred from the \textit{K}-band light, which was used
to model the LMXB distribution in \citet{Voss2}. The observed
\textit{K}-band luminosity of the region is 4.4$\cdot 10^{10}
L_{\odot,K}$, and integrating over the isochrone, 
we find a mass-to-light ratio of $(M/L)_K$=0.76, giving a total
mass of 3.4$\cdot  10^{10} M_{\odot}$, compatible with the 
\textit{R}-band estimate.
As a consistency check, we estimate the mass, using the mass-to-light
ratios of \citet{Bell}. With the bulge colour 
(\textit{B}-\textit{V})=0.95 \citep{Walterbos}, 
we find a bulge mass of 3.75 and 
3.73 $\cdot 10^{10} M_{\odot}$ from the \textit{R}-band and the
\textit{K}-band, respectively.

In figure \ref{fig:surplus} the $\rho_*^2$ profile, integrated over
the line of sight, is compared to the observed distribution of
surplus sources, which was calculated by subtracting the best-fit
model of LMXBs and CXBs from the observed radial distribution of X-ray
sources  (section \ref{sect:data}, figure \ref{fig:fit}). 
It is obvious that the distributions agree well
everywhere outside $\sim$ 4 arcsec.
In the innermost 4 arcsec of M31 the mass model of
\citet{Riffeser} features a sharp increase in density, absent in the 
distribution of X-ray sources. 
This increase is due to a stellar disc of high density surrounding the
central super-massive black hole \citep{Bender}.
In this paper we do not try to model the environment in this region
and exclude the disc component.
The stellar model used for computation of the encounter rates  is
described by the following distribution: 
\begin{equation}
\rho_{bulge} =\rho_0 10^{-0.4(7.1a_{bulge}^{1/4}+0.61)}
\end{equation}
where
\begin{equation}
a_{bulge}=\frac{0.254z_0^2+\sqrt{0.254^2z_0^4+4(x_0^2+y_0^2+1.11)z_0^2}}
{2}
\end{equation}
with $a_{bulge}$, $x_0$, $y_0$ and $z_0$ expressed in arcmin. The
inclination of the bulge coordinate system is assumed to be $77^\circ$,
and $\rho_0=4.34\cdot10^{4} M_{\odot}$ pc$^{-3}$ (using our mass
to light ratio $(M/L)_R=3.27$). This gives a bulge mass (within
12 arcmin from the centre) of $3.87\cdot 10^{10} M_{\odot}$ 
and
\begin{equation}
\int\rho_*^2 dV= 4.6\cdot 10^{11} M_{\odot}^2 ~{\rm pc}^{-3}
\nonumber
\end{equation}


It is now straightforward to calculate the rates of tidal captures
and collisions. Following the equations of section \ref{sect:intro} the 
rates are given by
\begin{equation}
R_{M31}=\int_{bulge}\left(\frac{\rho_{\ast}}{<M>}\right)^2 dV\cdot f_{ns}\int_{M_{low}}^{M_{high}}
f(M)
\gamma dM
\end{equation}
where $M_{low}-M_{high}$ is the initial mass range for the type of
stars for which the rates are calculated. 
The rates for different types of encounters  are summarized in Table
\ref{tab:M31rates}. For clarity the channels expected to
lead to the formation of LMXBs are written in bold font.

\subsubsection{Numbers of X-ray sources} 

We turn now to the numbers of  of dynamically formed X-ray sources. 
As it is obvious from Table \ref{tab:M31rates} (column 4), the
number of ``normal'' presistent LMXBs with a neutron star accreting
from a main sequence companion $M_*>0.3M_\odot$, which constitute
the majority of the primordial LMXBs, 
is negligibly small (BH capture products with $M_*>0.3M_\odot$  
donors are expected to be transients and are discussed below). 
The two main contributions to the population of dynamically formed
sources come from the tidal captures of black holes and neutron stars
by very low mass MS stars, and from  collisions of compact objects
with RGB stars (columns 3 and 5). 
In computing the numbers of sources from equation \ref{eq:nx} we take into 
account that the 
the evolutionary timescales of all types of dynamically formed X-ray
sources are much shorter than the life time of the bulge. 
Therefore the systems formed via tidal capture by $M_*>0.3M_\odot$
stars will pass through the phase of the very low mass companion in
the  end of their life time, adding to the numbers of persistent and
transient sources of this type.  Similarly, a capture product of, for
example,  a $0.3M_\odot$ star will go through the transient phase in
the beginning of its X-ray active phase  and will become a persistent
source after the donor star is depleted below $\sim 0.10-0.15 M_\odot$. 
We thus predict $\sim 24$ and $\sim 5$ persistent X-ray sources with
black holes and neutron stars respectively, accreting from the very
low mass stars. To this number should be added the number of
ultra-compact X-ray binaries produced via collisions of compact
objects with red giants, which is $\sim 3$. The total number of
predicted persistent sources is compatible with, albeit somewhat
larger than the observed number of surplus sources, $\sim 21$. Given
the number and magnitude of uncertainties involved
in the calculations and the simplifications made, we consider this as a
good agreement.  

Based on the range of the donor masses corresponding to unstable mass
transfer (figure \ref{fig:mdot}), we predict $\sim 30$ BH and $\sim 22$
NS transient  sources with very low mass donors $M_*<0.3 M_{\odot}$,
as well as $\sim 3$ BH transient sources with MS donors $>0.3
M_{\odot}$. Furthermore, 
exchange reactions might contribute with a number of LMXBs with RGB
donor stars, that are also transient, but duration of their active
phase is restricted by the life time of the red giant
donor. 
The number of transients observed at any given moment in time depends
on their duty cycle. Taking Galactic black hole transients with the
main sequence donor as an example, one could expect a duty cycle of
$\sim 1/50$, giving one bright transient in $\sim 15$ years.
As for the transients with very low mass donors,
one can use the accreting msec pulsars as an example of NS
systems. SAXJ1808.4-3658 has  outbursts lasting for $\sim
2-3$ weeks every $\sim 2$ years, and the duty cycle is therefore
$\sim$0.03. Assuming crudely that it is the same for BH and NS systems,
we would expect 1.5 transient sources at any given time. 
The outbursts of accreting msec pulsars in our Galaxy are
characterized by low peak luminosities, $\log(L_X)\la
36-36.5$. Therefore many, if not most, of outbursts from these sources
will be missed in a Chandra survey of the type reported in
\citet{Voss2} which detects mostly brighter transients, with the peak
luminosity of $\log(L_X)\ga 36.5$. This explains why \citet{Voss2}
have not found any excees in the number of transient sources close
to the galactic center -- the fraction of transients detected inside
1 arcmin from the center (5 out of 28 in 29 Chandra observations
with the time span of  $\sim 5$ years) agrees with the fraction of
stellar mass contained in this region. 
On the other hand Chandra observations of our Galactic Center, having
much better sensitivity, have indeed revealed overabundance of faint
transients \citep{Muno}.

\begin{table*}
\begin{center}
\begin{tabular}{llcccccc}
\hline
Metallicity & Type & MS($<0.3M_{\odot}$) & MS($>0.3M_{\odot}$) & RGB & HB &
AGB & Exchange\\
\hline
0.2 solar & Tidal capture & {\bf 8.5} (10.4) & {\bf 29.3} (32.5) & 7.0 & - & - & 203.4\\
& Collisions & 5.6 & 56.1 & {\bf 24.3-27.7}  & 4.6 & {\bf 0.4-1.1}  & {\bf 15.3}\\
\hline
0.02 solar & Tidal capture & {\bf 5.5} (6.6) & {\bf 17.3} (18.9) & 2.9 & - & - & 117.3\\
& Collisions & 3.6 & 31.6 & {\bf 10.1-11.7}& 2.0 & {\bf 0.3-0.6}& {\bf 8.8}\\
\hline
\end{tabular}
\caption{Total encounter rates for 140 Galactic globular clusters from
\citet{Harris} for which sufficient structural parameters are known,
calculated assuming metallicity of 0.2 and 0.02 solar. Entries in bold  
indicate paths expected to lead to the formation of LMXBs. The rates
are given in LMXBs/Gyr. 
These 140 GCs contain the 13 LMXBs observed in the Galactic GC system.
The notation is the same as in table \ref{tab:M31rates}}
\label{tab:GCall}
\end{center}
\end{table*}

\begin{table*}
\begin{center}
\begin{tabular}{lclcccccc}
\hline
Population & LMXBs observed &Type & MS($<0.3M_{\odot}$) & MS($>0.3M_{\odot}$)
& RGB & HB & AGB& Exchange\\
\hline
Red GCs & 8 & Tidal capture &  {\bf 2.7} (3.1) & {\bf 7.9} (8.7) & 1.6 & - & - & 53.3\\
& &Collisions & 1.8 & 18.4 & {\bf 7.9-9.2} & 1.5 & {\bf 0.1-0.4}  & {\bf 4.0}\\
\hline
Blue GCs & 5 & Tidal capture & {\bf 3.8} (4.6) & {\bf 12.6} (13.8) & 2.2 & - & - & 86.1\\
& &Collisions  & 2.4 & 21.3 & {\bf 6.8-7.8 } & 1.3 & {\bf 0.2-0.4} & {\bf 6.5}\\
\hline
\end{tabular}
\caption{Total encounter rates calculated separately for red and
the blue Galactic globular cluster subsystems, assuming metallicity
of 0.2 and 0.02 solar respectively. Bold entries 
indicate paths expected
to lead to formation of LMXBs, and the rates are given as LMXBs/Gyr.
The notation is the same as in table \ref{tab:M31rates}.}
\label{tab:GCdiv}
\end{center}
\end{table*}

\subsection{Globular Clusters}
\label{sect:gc}

Due to high efficiency of the mass segregation in globular clusters
the (retained) neutron stars will be much more centrally concentrated
than low mass stars. Assuming that stellar density and velocity
dispersion are approximately constant over the region occupied by the
neutron stars, one can approximately write:
\begin{equation}
\label{eq:GCrates}
\int_Vn_{ns}n_{\ast}\Gamma \,dV\simeq n_{\ast}\Gamma_c\int_Vn_{ns}\,dV=kN_{ns}\,n_{\ast}\Gamma_c
\end{equation}
where $N_{ns}$ is the total number of neutron stars in the globular
cluster under consideration, $n_*$ is the central density of stars,
$\Gamma_c$ is the central value of $\Gamma$ 
(equation \ref{eq:gamma_large})
and $k\la 1$ is a constant accounting for inaccuracy of this
approximation. 
Assuming that the distribution of normal stars follows the analytic
King model \citep{King}, and that the NSs are in thermal
equilibrium with the stars at turn-off (0.80-0.85 $M_{\odot}$)
\citep[as in][]{Lugger,Grindlay2},
we estimated that $k\sim0.2-0.3$. 
Total thermal equilibrium is generally not reached, the value of $k$ is
therefore slightly lower. In the following we will use a value of
$k=0.2$ for all globular clusters.

We use the catalogue of \citet{Harris} for the globular clusters
parameters required to estimate the formation rates of LMXBs
in the Galactic GCs. Of the 150 GCs included in the catalogue, the
parameters are missing for 10, and we ignore these. The stellar
populations in the GCs were modelled using the isochrones of
\citet{Girardi}, with an age of 11 Gyr \citep{Salaris}.
The $N_{ns}$ for each GC was computed as follows. 
Assuming the initial mass function of \citet{Kroupa}, we used the
integrated light of the isochrones to compute the present day
mass-to-light ratio and from the total \textit{V}-band luminosity of
the GCs computed the 
IMF normalization. Assuming further that all stars with the initial
mass in the range of $8-30\, M_\odot$ have become neutron stars 
and retention factor of 10 per cent \citep{Drukier,Pfahl} we finally compute
the present day number of the neutron stars in each globular cluster,
$N_{ns}$. 
On the other hand, we assumed that the present day mass function in
the GC centers is flat. With this mass function we
again use the integrated \textit{V}-band light of the isochrones to
calculate $n_{\ast}$ from the \textit{V}-band luminosity density $\rho_V$ 
given in \citet{Harris}.
For the 56 GCs in \citet{Pryor} we use their central velocity
dispersions $v_0$ needed to compute the encounter rates. 
The remainong GCs were dealt with as follows.
>From the virial theorem we expect that $v_0\sim Kr_c\sqrt{\rho_{0}}$,
where $r_c$ is the core radius of the GCs, and $\rho_{0}$ is the
central density; we further assumed that $\rho_V\propto\rho_{0}$.
We performed the least square fit to the known central velocity
dispersions in 56 GCs and found $K=0.18$ km s$^{-1}$ and 
$0.17$ km s$^{-1}$ for the metal-rich and metal-poor  GCs respectively 
(assuming that  $r_c$ is in pc and $\rho_V$ in $M_{\odot,V}$
pc$^{-3}$). These values have been used to find $v_0$ for the
remaining 84 GCs.

\subsubsection{Metallicity effects}

In order to study the metallicity dependence of the encounter rates,
we compute the cumulated rates for two metallicities, 20 per cent,
and 2 per cent of the solar value 
(files {\tt isocz004.dat} and {\tt isocz0004.dat} from
\citet{Girardi}) which are representative of the red and blue GC
populations, respectively. The results are presented in 
table \ref{tab:GCall} and show a $\sim$1.5-2.5 increase in the
encounter rates for the higher metallicity case.

The metallicity dependence in our calculations is 
mainly due to two factors. (1) As noted by \citet{Bellazzini} the radii of
metal-rich stars are larger, and therefore the rates of tidal captures
and collisions are higher. Furthermore the duration of the RG phase
is longer for metal-rich stars. As demonstrated by \citet{Maccarone3}
this effect can maximally lead to an enhancement of the
cross-sections and rates by $\lesssim$
60 per cent, and most likely $\sim$ 30 per cent. Our results are
consistent with this, showing a $\sim$ 20 per cent increase in
tidal captures by MS stars ($>0.3 M_{\odot}$) and $\sim$ 50 
per cent increase in collisions with RGB/AGB stars. For exchange
reactions the effect is negligible. 
(2) Theoretical isochrones predict that the \textit{V}-band
mass-to-light ratio of the metal-rich population is higher than that
of the metal-poor population. As the stellar
densities are given in  \citet{Harris} in the form of \textit{V}-band
luminosity density, the encounter rate is proportional to
$\rho_*^2\propto (M/L)^2$. This could result in an additional $\sim
60$ per cent increase in the rates.   It is however unclear, whether
this is the 
case for real globular clusters -- observations indicate that the the
central mass-to-light ratio might be independent on the metallicity  
\citep{McLaughlin}. This could be due to the fact that the 
red GCs typically are more dynamically evolved (but not older)
than the blue ones and therefore have a flatter mass function
in their cores \citep{McClure,Vesperini,Piotto}. 
Moreover, these structural differences may be the true reason for the 
observed metallicity dependence of the abundance of dynamically
created sources in globular clusters as also noted by \citet{Bregman}.  

Thus our calculations do indicate a moderate metallicity
dependence of the encounter rates. It is however insufficient to
explain observations. Indeed, there are $\sim$3 times as many LMXBs in
red GCs as in blue GCs of the same size in the Galaxy \citep{Grindlay,
Bellazzini}, where 8 out  of 13 LMXBs are observed in the red GC
system containing 46 out of the total number of 140 GCs with known
metallicities (assuming a division at [Fe/H]=-1). Similar trend is
observed in in other galaxies \citep{Kundu, Sarazin,Kim}.

\subsubsection{Predicted rates and numbers of X-ray sources}

To predict the total rates of LMXBs formation in the Galactic
GCs, we divide the GCs into two subpopulations depending on
metallicity, red (46 GCs) and blue (94 GCs). The cumulative rates for
these two subpopulations are
then calculated as above, assuming all red GCs to have 0.2 
solar metallicity and all blue ones to have 0.02 solar
metallicity. The results are given in table \ref{tab:GCdiv}.
As it can be expected from figure \ref{rates1}, all three processes
give comparable contributions.

For metal-rich clusters, these rates predict $\sim 1.2$ X-ray
binaries with the companion mass $>0.3 M_\odot$ due to tidal captures, 
with an additional $0.5-1.0$ such binaries from exchange reactions, 
$\sim 1.5$ UCXBs and
$\sim 3$ fainter LMXBs  with very low mass companion. 
Corresponding to  $\sim 6$ sources overall, this is in a good
agreement with the total number of LMXBs observed in metal-rich
clusters (8).  On the other hand, we do overpredict the numbers of
X-rays sources in the metal-poor GCs by a factor of $\sim 1.5$ -- 
although our calculations do show the expected metallicity dependence, 
it is compensated by the larger number of metal-poor clusters. 
Note that the number of bright sources with $M_d>0.3M_\odot$ main
sequence companion dependes critically on the rate of magnetic
braking. The above numbers have been computed with the standard
prescription of \citet{Rappaport}. The weaker variants of magnetic
braking 
\citep[e.g.][]{Ivanova_Taam} may give upto a factor $\sim 5-10$ longer
lifetimes and, consequently, larger numbers of LMXBs with  $M_d>0.3
M_\odot$ donors. This can change the overall numbers for globular
clusters, but is insignificant factor in the M31 bulge calculations,
due to negligible contribution of these systems  there. 

It is interesting to compare the numbers of ultra-compact systems.
Considering metal rich clusters only, 2 of the 8 LMXBs have measured
orbital periods 
$\lesssim$ 1 h and are therefore most likely UCXBs \citep{Benacquista}.  
Of the 6 others 4 have undetermined periods and could
therefore be either UCXBs or traditional LMXBs. The final 2 have
orbital periods $>5$ h. Thus, there may be from 2 to 6 short period
systems. We predict $\sim 1.5$ UCXBs formed in the
collisions with red giants. In addition, the LMXBs with the very low
mass donor stars, $M_d\la 0.15M_\odot$, for which the predicted number
is $\sim 3$, will also have short orbital periods and faint optical
counterparts and may contribute to the observed statistics of UCXBs,
giving a prediction of $\sim 4.5$ short period systems in total.  


\begin{table}
\begin{center}
\begin{tabular}{ccccc}
\hline
&47 Tuc & & $\omega$ Cen\\
&Tidal & RG-NS & Tidal & RG-NS\\
\hline
Our study& 4 & 2 & 11 & 3\\
\citet{DaviesB}& 3 & 1 & 14 & 2\\
\hline
\end{tabular}
\caption{Comparison of the LMXB production rates for two Galactic GCs
with the results of \citet{DaviesB}. When computing our numbers we
adjusted the parameters of the stellar environment according to the
assumption of \citet{DaviesB}, as described in the text.
The rates are given in units of Gyr$^{-1}$.
}
\label{tab:Dav}
\end{center}
\end{table}

\subsection{Comparison with previous studies}

As we have already emphasized above, there is only a handful
of studies dedicated the quantitative predictions of the
formation rates and numbers of LMXBs in real globular clusters.

The results of our globular cluster calculations  agree with the
estimates of \citet{Verbunt2003} of  the relative production rates of
LMXBs in different Galactic GCs. Later on this method has been used to 
succesfully explain the observed distribution of X-ray sources over
Galactic \citep{Pooley} as well as for extra-Galactic GCs
\citep{Sivakoff}. 

One of the most detailed investigations so far has been done by
\citet{DaviesB} who considered LMXB formation in realistic
models of $\omega$ Cen and 47 Tuc globular clusters. 
When compared blindly, their results appear to differ from our
calculations  for the same two clusters. However, this is due to different
assumptions on the stellar environments used in their study.
The main differences are, that in \citet{DaviesB}: (i) the NS
depletion effect due to supernova kicks was not taken into  account
(i.e. 100\% NS retention factor has been assumed), 
(ii) it was assumed that stars more massive than 6$M_{\odot}$ produce
NSs as compared with  8$M_{\odot}$ boundary used in this paper and 
(iii) the initial mass function for $\omega$ Cen was much flatter than
Kroupa IMF used in this paper. 
In order to compare with their results, we modified our
calculations to be consistent with these assumptions and we found
good agreement between the two studies, as demonstrated by the table
\ref{tab:Dav}.  We note that there also is a number of more subtle
differences, not taken into account here, which may explain the
remaning differences. 
Finally, consistent with our results, they find that the production 
rate due to exchange
reactions is similar to the rate from the other two channels.

Our calculations are also consistent, within a factor of 2, with
those of \citet{IvanovaU}, who studied the rate of collisions 
between RGs and NSs in a small sample of Galactic GCs. 

The only investigation of the formation of LMXBs in galactic bulges is  
the study by \citet{LeeN} of tidal captures near the Galactic centre.  
Extrapolating from calculations of captures by stars of
0.5 $M_{\odot}$ only, they found that while tidal captures
can happen in numbers there, the majority of
these would actually be collisions. This is consistent with
our results for encounters with stars of this mass.

\section{Conclusions}
\label{sect:conclusions}

\begin{figure}
\includegraphics[width=\hsize]{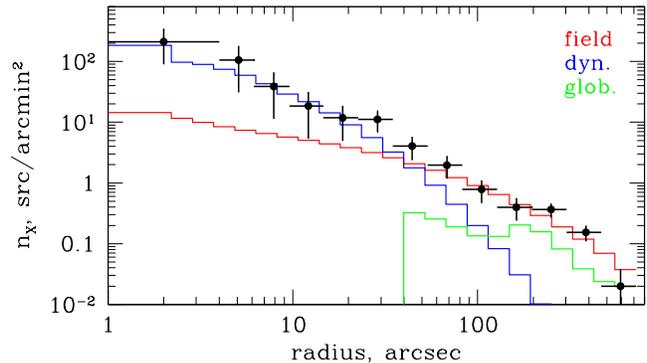}
\caption{The observed radial distribution of the X-ray sources in the
bulge of M31, compared with the expected contributions  of
different sub-populations of low-mass X-ray binaries:
primordial (red, the broadest of the three histograms), binaries in
globular clusters (green, with zero source density in the center) and
binaries  formed through dynamical interactions in the inner bulge of
M31 (blue, the most peaked distribution).   
The total numbers of sources are summarized in table
\ref{tab:numbers}.}
\label{fig:all}
\end{figure}

\begin{table}
\begin{center}
\begin{tabular}{lc}
\hline
Type & Number\\
\hline
Background sources & 29\\
Primordial LMXBs & 64\\
LMXBs in globular clusters & 21\\
LMXBs dynamically formed in the bulge  & 21\\
\hline
\end{tabular}
\caption{Numbers of X-ray sources of different origin in the 
bulge of M31, $r<12$ arcmin, $L_X>10^{36}$ erg/s}
\label{tab:numbers}
\end{center}
\end{table}

We have studied the spatial distribution of the luminous X-ray 
point sources ($L_x>10^{36}$ erg s$^{-1}$) in the bulge of M31. We
show that there is a significant increase in the specific frequency of 
sources, per unit stellar mass, in the inner $\approx 1$ arcmin. This
behaviour is similar, although smaller in the magnitude, to that
observed in globular clusters. The radial distribution of the surplus
sources follows the $\rho_*^2$ profile. All these suggest that the
surplus sources are dynamically created in stellar encounters in the
high stellar density environment of the inner bulge of M31.
This is further confirmed by the peculiarity of their luminosity
distribution, which resembles that of the globular cluster sources in
M31 and our Galaxy \citep{Voss2}.

It has long been known that dynamical interactions are responsible
for the relatively large number of X-ray sources observed in globular
clusters, but this is the first evidence of the dynamical formation of
LMXBs  in the vicinity of a galactic center. The stellar velocities in bulges
are higher than in globular clusters by a factor of $\sim 5-10$. We
therefore performed a detailed study of the velocity dependence of the
three main dynamical processes  leading to the formation of LMXBs:
tidal captures of a compact objects by main sequence stars, collisions
between evolved stars and compact objects and the  exchange of a
compact object into an already existing binary. Another major factor
affecting 
the overal encounter rates and the numbers of dynamically formed LMXBs
is the high efficiency of the mass segregation in globular clusters,
which modifies significantly the spatial distributions of objects of
different mass and affects the present day mass function in different
parts of a globular cluster. In addition, due to the relative
shallowness of the potential well, the populations of compact objects
are significantly depleted in globular clusters. 

We found that while exchange reactions are potentially the dominant
formation channel in globular clusters (although stellar collisions
might decrease 
the importance of this channel significantly), this process is
relatively unimportant in M31. Similarly, tidal captures of NSs by
main sequence stars of mass $>0.3 M_{\odot}$ are important in globular
clusters, but not in M31. Instead the main formation channel
is tidal captures of compact 
objects by low mass ($<0.3 M_{\odot}$) stars, with some contribution
from collisions between red giants and compact objects. While
the geometrical collision rate is high enough to explain the total
number of sources from the latter channel, the majority of the collisions
are unlikely to lead to the formation of a binary system, as the
binding energy of the envelopes of most RGB/AGB stars is too low to
capture a compact object in a high velocity environment. 
We conclude that the majority of the sources in M31 are short-period
binaries, and in contrast to globular clusters many of them have BH
accretors. We note that the BH binaries with a very low mass
companion may become persistent X-ray sources after the donor star is
depleted below, $M_d\la 0.15M_\odot$, due to small size of the
accretion disc and the positive dependence 
of the gravitational breaking rate on the mass of the primary.
We also predict for M31 a large number of faint transients,
similar to the accreting msec pulsars in our Galaxy. Overall, we have
been able to explain the spatial distribution and absolute numbers of
surplus sources detected in the inner bulge of M31 as a well as the
statistics of LMXBs in the metal rich globular clusters. However, we
overpredict by a factor of $\sim 1.5$  the population of LMXBs in the
metal poor clusters.  

Finally, the sub-populations of low-mass X-ray binaries in the bulge
of M31 are summarized in Fig.\ref{fig:all} and Table
\ref{tab:numbers}. 

\bigskip
{\em Aknowledgements} The authors would like to thank Natasha Ivanova
for numerous discussions and useful comments on the initial version
of the manuscript. We also thank our referee, Simon Portegies-Zwart,
for helpful and  constructive comments.

\bsp

\label{lastpage}
\end{document}